\documentclass[aps,
  prd,
  reprint,
  superscriptaddress,
  nofootinbib
]{revtex4-2}

\usepackage[dvipdfmx]{graphicx}
\usepackage{graphicx,bm,color}
\usepackage{amsmath,amssymb}
\usepackage{hyperref}
\usepackage{amsfonts}
\usepackage{cases}
\usepackage{cancel}
\usepackage{ulem}
\usepackage{makecell}

\newcommand{\gsim}{\lower.7ex\hbox{$\;\stackrel{\textstyle>}{\sim}\;$}}
\newcommand{\lsim}{\lower.7ex\hbox{$\;\stackrel{\textstyle<}{\sim}\;$}}

\begin{document}

\title{Modular TM$_1$ mixing in light of precision measurement in JUNO}

\author{Wen-Hao Jiang}
\email{jiangwenhao25@mails.ucas.ac.cn}
\affiliation{School of Fundamental Physics and Mathematical Sciences,
Hangzhou Institute for Advanced Study, UCAS, Hangzhou 310024, China}
\affiliation{Institute of Theoretical Physics,
Chinese Academy of Sciences, Beijing 100190, China}
\affiliation{University of Chinese Academy of Sciences,
Beijing 100049, China}

\author{Ruiwen Ouyang}
\email{ruiwen.ouyang@ucas.ac.cn}
\affiliation{School of Fundamental Physics and Mathematical Sciences,
Hangzhou Institute for Advanced Study, UCAS, Hangzhou 310024, China}

\author{Ye-Ling Zhou}
\email{zhouyeling@ucas.ac.cn}
\affiliation{School of Fundamental Physics and Mathematical Sciences,
Hangzhou Institute for Advanced Study, UCAS, Hangzhou 310024, China}

\begin{abstract}

This paper investigates the landscape of models based on modular $S_4$ symmetry that predicts the trimaximal TM$_1$ mixing pattern for leptonic flavor mixing, and explores their parameter spaces with constraints from the latest high-precision measurement on $\theta_{12}$ and $\Delta m^2_{21}$ given by JUNO experiment. We review on how the mixing pattern arises from residual symmetries after the spontaneous breaking of a flavor symmetry, via an appropriate vacuum alignment of modular fields and flavon fields. We show three different models that realize the TM$_1$ in three approaches with the same symmetry structure. Due to different model building strategies used, predictions on the relations between the lightest neutrino mass and the effective Majorana mass in neutrinoless double beta decay are different, making them distinguishable.



\end{abstract}

\maketitle


\section{Introduction}
Leptonic flavor mixing remains a mystery in particle physics. 
Following decades of sustained and systematic efforts, three mixing angles and two mass-squared differences have been measured \cite{ParticleDataGroup:2024cfk}. The coming goals are the determination of neutrino mass ordering (normal $m_1 < m_2 < m_3$ or inverted $m_3 < m_1 < m_2$) and measurement of the Dirac phase $\delta$. 
JUNO \cite{An:2015jdp}, Hyper-K \cite{Acciarri:2015uup}, and DUNE \cite{Abe:2018uyc}, the three representative large-scale neutrino oscillation experiments of this stage, aim to achieve these goals. 
Meanwhile, measurements of other oscillation parameters with precision at sub-percent level will be performed. 
After less than two months of running, JUNO released the first data on the measurement of $\Delta m^2_{21}$ and $\sin^2\theta_{12}$ with best-fit values $\pm 1\sigma$ given by \cite{JUNO:2025gmd}
\begin{align}
\sin^2 \theta_{12} &= 0.3092 \pm 0.0087 \, , \nonumber \\
\Delta m^2_{21} &= (7.50 \pm 0.12) \times 10^{-5} \text{eV}^2\,, 
\end{align}
for the normal mass ordering scenario, which has already shown a better precision than the previous global fit for all past measurements~\cite{Esteban:2024eli}.

To explain the observed neutrino oscillation parameters, various of simple mixing ansatzes for the lepton flavor mixing were proposed. Due to their simplicity in describing experimental data and predictability from flavor symmetries, they have been widely studied in neutrino theories (see, e.g., reviews \cite{King:2017guk,Ding:2024ozt,Ding:2023htn,Feruglio:2019ybq,Xing:2020ijf} and a recent paper~\cite{Xing:2025bdm}). 
A particular example is a constant mixing pattern called tri-bimaximal (TBM) mixing \cite{Harrison:2002er,Xing:2002sw}, although it has been excluded by the observation of a relatively large $\theta_{13}$ in reactor neutrino experiments \cite{DayaBay:2012fng,RENO:2012mkc}. 
Among various of pre-existing simple mixing patterns in the literature, the trimaximal TM$_1$ mixing (proposed by~\cite{Xing:2006ms} and further discussed in~\cite{Lam:2006wm,Albright:2008rp,Albright:2010ap,Grimus:2008tt,He:2011gb}), a partially constant mixing which inherits the first column of the TBM form, offers an elegant description for the observed lepton mixing.

For better understanding the trimaximal TM$_1$ mixing pattern, non-Abelian discrete symmetries were introduced as origins of flavor mixing, and a particular mixing pattern arises from the spontaneous breaking of flavor symmetry. 
For example, the TBM pattern can be realized from the permutation group $S_4$. It is enforced if the charged lepton mass term keep a $Z_3^T$ symmetry and the neutrino mass term is invariant under $Z_2^S$ and $Z_2^U$ symmetries, all of which are residual symmetries after $S_4$ breaking \cite{Lam:2008rs}\footnote{$T$, $S$ and $U$ are generators of $S_4$ satisfying $T^3 = S^2=U^2  = (SU)^2 = (TU)^2 = (ST)^3 = (STU)^4 = 1$.}. Relaxing the residual symmetry in the neutrino sector to be $Z_2^{SU}$ gives rise to TM$_1$ mixing \cite{deMedeirosVarzielas:2012apl,Luhn:2013vna}. In the traditional flavor symmetry approach, the symmetry breaking is achieved via vacuum expectation values (VEVs) of a series of scalar fields which are called flavons, and the flavor texture succeeds the special direction of the VEV alignment.

Modular symmetries, as a non-linearly realized approach, provide an alternative explanation that the flavor texture might arise from modular forms of a modular symmetry \cite{Feruglio:2017spp}. Here, a modular form is a holomorphic function of the modular field $\tau$, and the symmetry breaking as well as the special pattern of modular forms are achieved via the VEV of the modular field. In the minimal setup (e.g.~\cite{Penedo:2018nmg,Novichkov:2018ovf}), it only needs a modular field, which has only two real degrees of freedom, acquiring a VEV, without the necessity of introducing many flavon fields aligned in some specific ways (e.g.~\cite{King:2013eh}), to achieve the desired phenomenology. Thus, this approach reduces physical degrees of freedom very efficiently.

\begin{figure*}[t]
    \centering
    \includegraphics[width=0.9\linewidth]{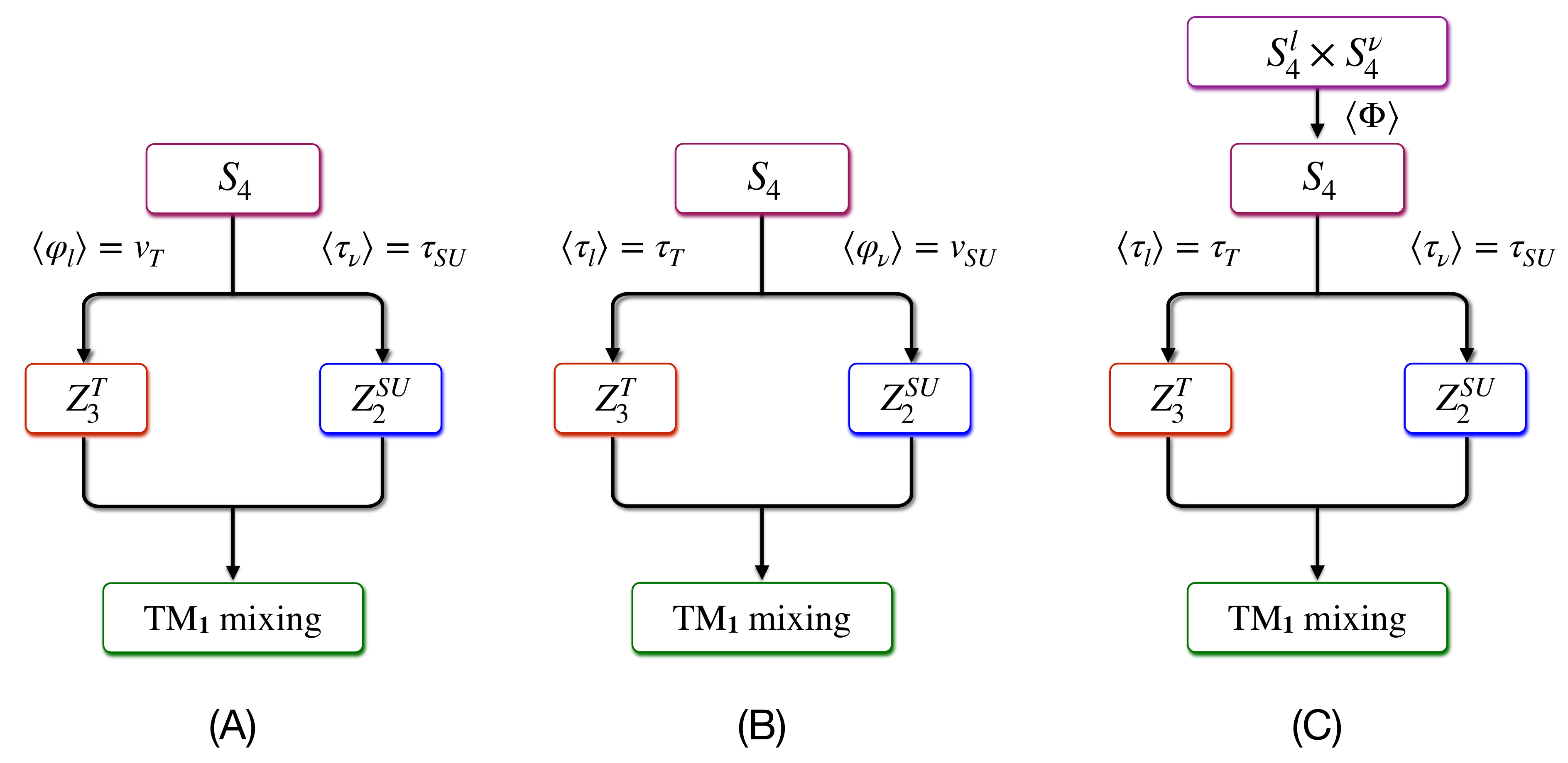}
    \caption{A few approaches to realise TM$_1$ mixing in modular flavor models. Here $\varphi_l$, $\varphi_\nu$ denote flavons associated with the charged-lepton and neutrino sectors, respectively, while $\tau_l$, $\tau_\nu$ denote modular fields.}
    \label{fig:approaches}
\end{figure*}

A modular field may gain the VEV at a fixed point, which is invariant under certain modular transformations. The lepton mass matrix at the fixed point preserves a residual symmetry as a subgroup of the modular symmetry. Implications of residual modular symmetries were suggested in the derivation of simple structures of modular forms \cite{Novichkov:2018ovf}. 
However, the minimal setup is not enough to accommodate the observed lepton mixing because a single modular field cannot achieve the misalignment of different residual symmetries in the charged lepton sector and neutrino sector \cite{Novichkov:2018yse}. 
Necessary extensions must be considered in realistic models. For example, by introducing a minimal set of flavons in the modular symmetry, the modular symmetry can be broken either in the charged lepton sector or in the neutrino sector via the flavon and modular field \cite{Kobayashi:2018vbk}. 
Another approach is to include distinguishable modular fields in the framework of multiple modular symmetries \cite{deMedeirosVarzielas:2019cyj}. This approach allows different residual symmetries from fixed points, e.g., $\tau_l = - \frac{1}{2} + i \frac{\sqrt{3}}{2}$ respecting $Z_3^T$ and $\tau_\nu = - \frac{1}{2} + \frac{i}{2}$ respecting $Z_2^{SU}$, predicting exactly TM$_1$ mixing \cite{King:2019vhv}.


The present paper aims to investigate the landscape of modular $S_4$ models that realize the ${\rm TM}_1$ mixing and explore their parameter spaces with constraints from the latest precision measurements in JUNO. We sketch three possible approaches with modular symmetries in Fig.~\ref{fig:approaches}. 
In these approaches, $\varphi_l$ and $\varphi_\nu$ denote the flavon fields associated with the charged-lepton and neutrino sectors, respectively. Their role is distinct from that of the modular fields $\tau_l$ and $\tau_\nu$, as the VEVs of $\varphi_l$ and $\varphi_\nu$ provide flavor symmetry breaking through vacuum alignment, whereas the VEV of $\tau$ fixes the values of the modular forms and, when located at a fixed point, determines the residual symmetry preserved in the corresponding sector. Therefore, in the realizations sketched in Fig.~\ref{fig:approaches}, different combinations of modular fields and flavons are considered to obtain the desired residual symmetries in both sectors which can realize the TM$_1$ mixing.

The rest of the paper is organized as follows. We begin with a review of the trimaximal ${\rm TM}_1$ mixing pattern in Section~\ref{sec:2}. We revisit the mathematical correlation between ${\rm TM}_1$ and residual symmetries  $Z_3^T$, $Z_2^{SU}$ of $S_4$, regardless of the traditional flavor symmetry approach or modular symmetry approach. 
This implies that one can take approaches in Figure~\ref{fig:approaches} to build concrete models featuring the ${\rm TM}_1$ mixing. In Section~\ref{sec:3}, we present explicit models with modular $S_4$ flavor symmetry, which feature ${\rm TM}_1$ but give different predictions on other observables. We perform numerical scans to fit these models in Section~\ref{sec:4}. The summary and conclusions are presented in Section~\ref{sec:5}.

\section{Trimaximal TM$_1$ mixing}\label{sec:2}
\subsection{TM$_1$ mixing in light of recent data}\label{sec:2.1}
\begin{figure*}[t]
    \centering
    \includegraphics[width=0.45\textwidth]{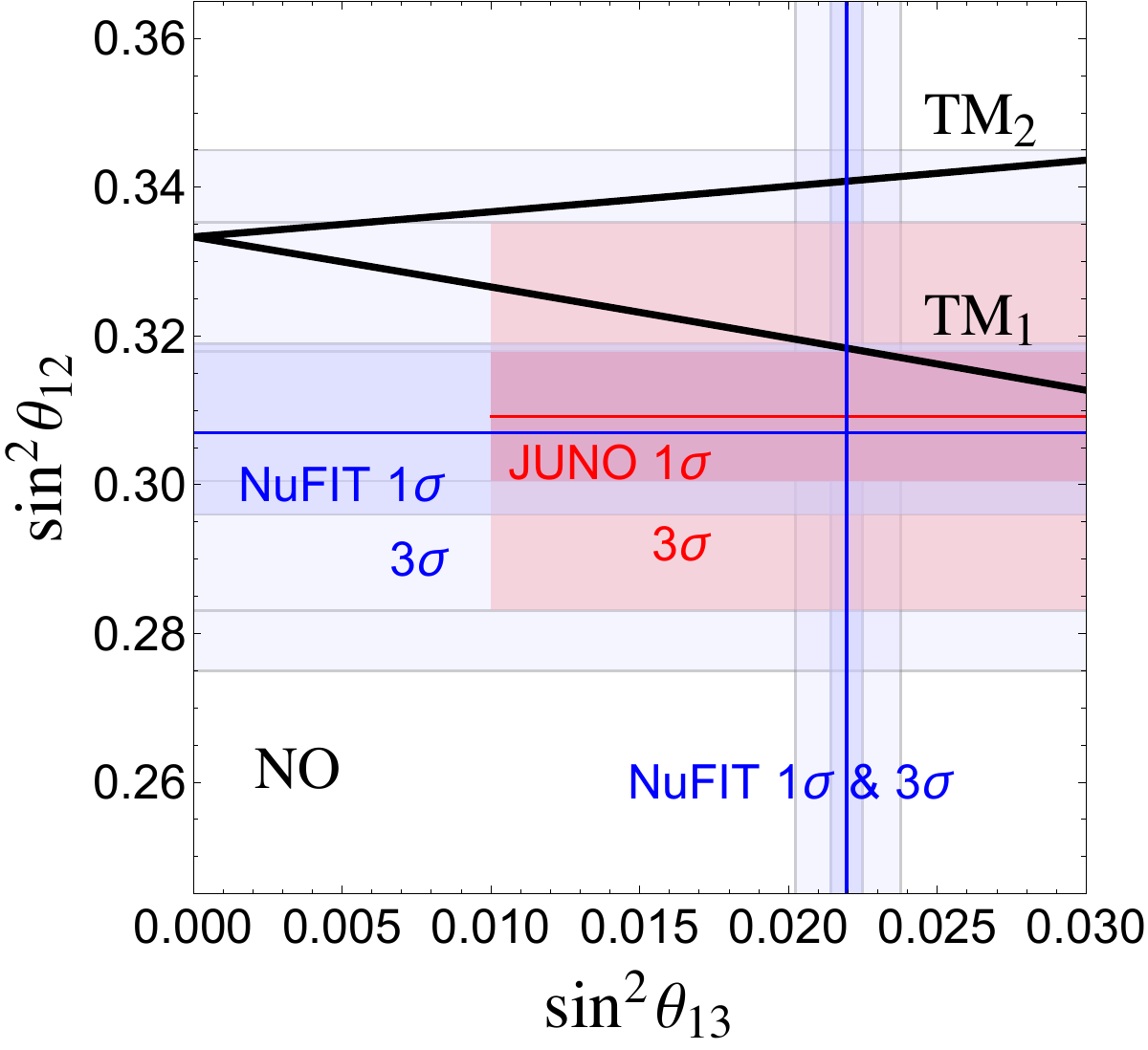}
    \hfill
    \includegraphics[width=0.45\textwidth]{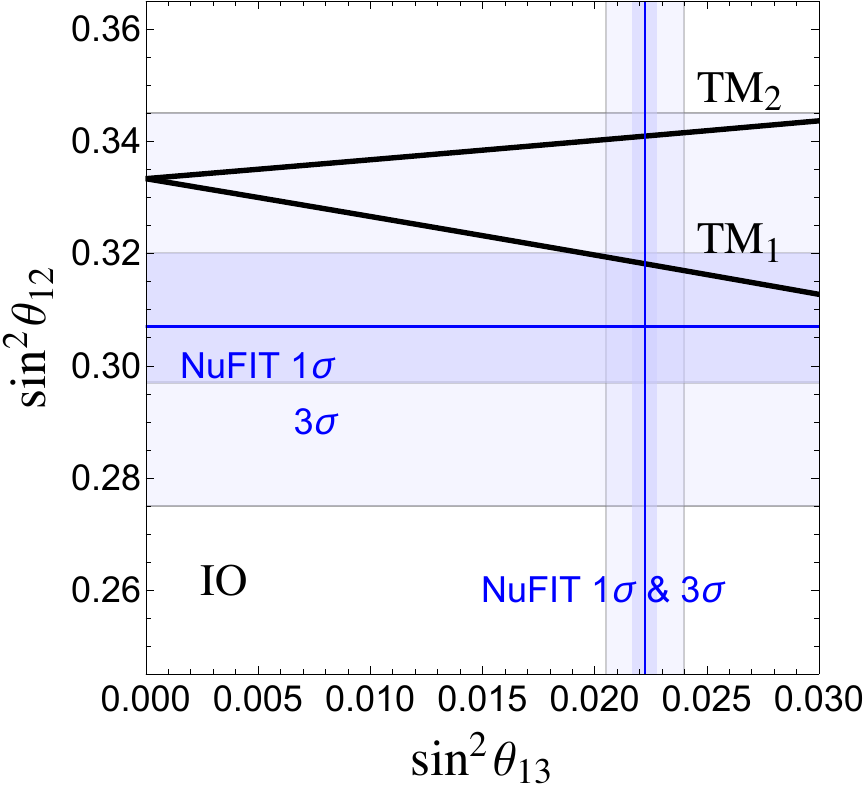}
    \caption{Sum rules between $\theta_{12}$ and $\theta_{13}$ in TM$_1$ and TM$_2$ mixing. Best-fits, $1\sigma$ and $3\sigma$ ranges from NuFIT 6.0 (blue)~\cite{Esteban:2024eli} and the first run of JUNO (red)~\cite{JUNO:2025gmd} are shown as comparison.}
    \label{fig:TM}
\end{figure*}

We begin by reviewing the trimaximal mixing used to explain the neutrino mixings in the literature. The PMNS matrix can be parameterized as~\cite{ParticleDataGroup:2024cfk}:
\begin{widetext}
\begin{equation}
U_{\rm PMNS}=\begin{pmatrix}
        c_{12}c_{13} & s_{12} c_{13} & s_{13}e^{-i\delta} \\ -s_{12}c_{23}-c_{12}s_{13}s_{23}e^{i\delta} & c_{12}c_{23}-s_{12}s_{13}s_{23}e^{i\delta} & c_{13 } s_{23} \\ s_{12}s_{23}-c_{12}s_{13}c_{23}e^{i\delta} & -c_{12}s_{23}-s_{12}s_{13}c_{23}e^{i\delta} & c_{13}c_{23}
    \end{pmatrix}
    \ 
    \begin{pmatrix}
        1 &0 & 0 \\ 0 & e^{i \frac{\alpha_{21}}{2}} & 0 \\ 0 & 0 & e^{i\frac{\alpha_{31}}{2}}
    \end{pmatrix} \, ,
\end{equation}
\end{widetext}
where $s_{ij} = \sin \theta_{ij}$, $c_{ij} = \cos \theta_{ij}$ for three mixing angles $\theta_{12}$, $\theta_{13}$, $\theta_{23}$, $\delta$ is the Dirac phase, and $\alpha_{21}$, $\alpha_{31}$ are irremovable Majorana phases for Majorana neutrinos. A particular mixing pattern, giving $s_{12}^2=1/3$, $s_{13}^2=0$, and $s_{23}^2=1/2$, is the so-called tri-bimaximal (TBM) mixing pattern~\cite{Harrison:2002er,Xing:2002sw}, with each entry, up to a phase difference, being constant,
\begin{eqnarray}
    \renewcommand{\arraystretch}{1.5}
    U_{\rm TBM} = \begin{pmatrix}
        \frac{2}{\sqrt{6}} & \frac{1}{\sqrt{3}} & 0 \\ \frac{-1}{\sqrt{6}} & \frac{1}{\sqrt{3}} & \frac{1}{\sqrt{2}} \\ \frac{-1}{\sqrt{6}} & \frac{1}{\sqrt{3}} & \frac{-1}{\sqrt{2}} 
        \label{eq:TBM}
    \end{pmatrix} \, .
\end{eqnarray}
This mixing ansatz is apparently excluded due to the nonzero mixing angle $\theta_{13}$, but still well approximates the observed values of $\theta_{12}$ and $\theta_{23}$ in the $3\sigma$ regimes. An extension of the TBM ansatz was thus proposed to incorporate the small but not tiny reactor angle $\theta_{13}$~\cite{Albright:2010ap}, where the lepton mixing matrices still preserve the first column of the TBM ansatz and leave the rest parameters unfilled, which is now called the trimaximal mixing ansatz, ${\rm TM}_1$, 
\begin{eqnarray}
    \renewcommand{\arraystretch}{1.5}
    U_{\rm TM_1} = \begin{pmatrix}
        \frac{2}{\sqrt{6}} & \times & \times \\ \frac{-1}{\sqrt{6}} & \times & \times \\ \frac{-1}{\sqrt{6}} & \times & \times 
    \end{pmatrix} \,    \label{eq:TM1}
\end{eqnarray}
up to phase differences, where the unfilled parameters can be fixed once the reactor angle $\theta_{13}$ is specified.  The TM$_1$ mixing leaves two sum rules between mixing angles and the CP phase,
\begin{eqnarray}
    {\rm TM_1:} && \quad  
    \sin^2 \theta_{12} = 1 - \frac{2}{3(1 - \sin^2 \theta_{13})} \, , \nonumber\\
    &&
    \quad  \theta_{23} \approx 45^\circ + \sqrt{2} \theta_{13} \cos \delta \,,
    \label{eq:sumrule}
\end{eqnarray}
The first sum in the above equation is enforced by the constant absolute value $|(U_{\rm TM_1})_{e1}| = 2/\sqrt6$, and the second sum rule is derived perturbatively for small $\theta_{13} $ after a $2$-$3$ unitary matrix taken into account \cite{King:2013eh}. 
Another widely studied mixing pattern is the ${\rm TM}_2$ mixing, which preserves the second column of TBM, 
\begin{eqnarray}
    \renewcommand{\arraystretch}{1.5}
    U_{\rm TM_2} = \begin{pmatrix}
        \times & \frac{1}{\sqrt{3}} & \times \\ \times & \frac{1}{\sqrt{3}} & \times \\ \times & \frac{1}{\sqrt{3}} & \times 
    \end{pmatrix} \, ,
    \label{eq:TM2}
\end{eqnarray}
up to phase differences, and predicts sum rules between mixing angles and CP phase \cite{King:2013eh}
\begin{eqnarray}
    {\rm TM_2:} && \quad  
    \sin^2 \theta_{12} = \frac{1}{3(1 - \sin^2 \theta_{13})} \, , \nonumber\\
    &&
    \quad  \theta_{23} \approx 45^\circ + \frac{1}{\sqrt{2}} \theta_{13} \cos \delta \,.
    \label{eq:sumrule_TM2}
\end{eqnarray}

Correlations between $\theta_{12}$ and $\theta_{13}$ in both TM$_1$ and TM$_2$ are presented in Fig.~\ref{fig:TM}. Given the values of $\theta_{13}$ from reactor experiments, the trimaximal ${\rm TM}_2$ mixing ansatz predicts $\theta_{12}\approx 36^{\circ}$, which is not consistent with JUNO data in $3\sigma$ region, as seen in the figure, even including the renormalization group (RG) effect running from a very high energy scale of the Standard Model~\cite{Zhang:2025jnn}.
On the other hand, ${\rm TM}_1$ predicts $\theta_{12}\approx 34^{\circ} $ for $\theta_{13}\approx 8.5^{\circ}$. This marginally touches on the $1\sigma$ boundary of JUNO data, where $\sin^2\theta_{12}=0.309\pm 0.009$~\cite{JUNO:2025gmd}, as shown in the Figure~\ref{fig:TM}. It can fit the JUNO data better if the small RG running effect is included~\cite{Zhang:2025jnn}. 
Therefore, in the following sections, we will mainly focus on the ${\rm TM}_1$ mixing. For the sum rules between $\theta_{23}$ and the CP phase $\delta$ in light of the recent data, we refer to reference~\cite{Ge:2025csr}.

\subsection{TM$_1$ mixing predicted in flavor symmetries}\label{sec:2.2}
Three different approaches presented in Fig.~\ref{fig:approaches} show the connection between those discrete residual symmetries and the mixing pattern of the PMNS matrix. In this section, we show a general symmetry-based deduction that the same mixing pattern can be achieved regardless of the different approaches taken. Before turning to realistic models, we give a brief review on how leptonic mixing arises from residual symmetries of a flavor symmetry and in particular, how TM$_1$ mixing arises from residual symmetries of $S_4$. 

Let us start by considering the complete Lagrangian at a high energy scale that is invariant under a flavor symmetry $G_f$, which is spontaneously broken by either the VEV of a flavon field in the traditional flavor symmetry approach, or by the VEV of a modular field in the modular symmetry approach. In both approaches, the residual symmetries after symmetry breaking, denoted as  $G_\ell$ and $G_\nu$ respectively in the charged lepton and neutrino sectors, are usually subgroups of $G_f$~\footnote{In a realistic model, sometimes the residual symmetries also include some additional accidental symmetries, but will not be considered in this work}.

At a low energy scale, the effective Lagrangian of charged lepton and neutrino mass terms with a spontaneously broken $G_{f}$ can be expressed as
\begin{eqnarray}
-\mathcal{L}_{\rm mass}=\overline{\ell_{ L}}M_{\ell}\ell_{R}+\frac{1}{2}\overline{\nu_{L}}M_{\nu}\nu_{L}^c+\text{h.c.}\, ,
\label{eq:Lleptons}
\end{eqnarray}

where it is assumed that neutrinos are Majorana particles. 

The residual symmetries $G_\ell$ and $G_\nu$ ensure that the operators $\overline{\ell_{L}}M_{\ell}\ell_{R}$ and $\overline{\nu_{L}}M_{\nu}\nu_{L}^c$ should be invariant under the transformation of $g_\ell \in G_\ell$ in the charged lepton sector and $g_\nu \in G_\nu$ in the neutrino sector, respectively. In the charged lepton sector, 
left-handed and right-handed charged leptons in their flavor space transform as
$\ell_{ L} \to \rho_{ L}(g_\ell) \ell_{ L}$, $\ell_{ R} \to \rho_{ R}(g_\ell) \ell_{ R}$ under $g_\ell$, leading to the mass term transforming as
$\overline{\ell_{L}}M_{\ell}\ell_{R} \to \overline{\ell_{L}} \; \rho_{ L}^\dagger(g_\ell) M_{\ell} \rho_{ R}(g_\ell) \;\ell_{R}$, where $\rho_{ L}(g_\ell)$ and $\rho_{ R}(g_\ell)$ are representation matrices of $g_\ell$ in the basis of $\ell_{ L}$ and  $\ell_{ R}$, respectively. 
The invariance of the mass term under $G_\ell$ thus implies
\begin{eqnarray}
 \rho_{  L}^\dagger(g_\ell) M_{\ell} \rho_{  R}(g_\ell) = M_{\ell} \,.
\end{eqnarray}
It is convenient to introduce a Hermitian matrix $H_\ell = M_\ell M_\ell^\dagger$ to avoid unphysical rotation in the flavor space of right-handed charged leptons, whose invariance implies
\begin{eqnarray} 
 \rho_{  L}^\dagger(g_\ell) H_\ell \rho_{  L}(g_\ell) = H_{\ell} \,.
\end{eqnarray}
This condition transmits to the restriction on the unitary matrix $U_\ell$ diagonalizing $H_\ell$, $U_\ell^\dagger H_\ell U_\ell = \hat{H}_\ell \equiv { \rm diag} \{ m_e^2, m_\mu^2, m_\tau^2 \}$, 
\begin{eqnarray} 
 \hat{H}_\ell \, [U_\ell^\dagger \rho_{ L}(g_\ell) U_\ell] = [U_\ell^\dagger \rho_{ L}(g_\ell) U_\ell] \, \hat{H}_{\ell} \,.
\end{eqnarray} 
In order to recover non-degenerate charged lepton masses, $U_\ell^\dagger \rho_{ L}(g_\ell) U_\ell$ must be diagonal 
\begin{eqnarray} \label{eq:residual_l}
U_\ell^\dagger \rho_{ L}(g_\ell) U_\ell = {\rm diag} \{ e^{i \alpha_1}, ~ e^{i \alpha_2}, ~ e^{i \alpha_3} \} \,.
\end{eqnarray}
Therefore, $G_\ell$ must be an Abelian symmetry and the values of phases $\alpha_i$ (for $i = 1,2,3$) depend on which residual symmetry is selected.

In the neutrino sector, the invariance of $\overline{\nu_{L}}M_{\nu}\nu_{L}^c$ under the transformation 
$\nu_{ L} \to \rho_{L}(g_\nu)\nu_{ L}$ requires
\begin{eqnarray}
 \rho_{ L}^\dag (g_\nu) M_{\nu} \rho^*_{ L}(g_\nu) = M_{\nu}
\end{eqnarray}
Similarly, given a unitary $U_\nu$ to diagonalize $M_\nu$, $U_\nu^\dag M_\nu U_\nu^* = \hat{M}_\nu \equiv {\rm diag} \{ m_1, m_2, m_3 \}$, 
The above equation is rewritten as 
\begin{eqnarray} 
 \hat{M}_{\nu} [U_\nu^\dag \rho_{ L}(g_\nu) U_\nu]^* = [U_\nu^\dag \rho_{ L}(g_\nu) U_\nu] \hat{M}_{\nu}
\end{eqnarray}
The condition requires $[U_\nu^\dag \rho_{ L}(g_\nu) U_\nu]$ to be real and diagonal, and thus the only choices could be 
\begin{eqnarray} \label{eq:residual_nu}
 [U_\nu^\dag \rho_{ L}(g_\nu) U_\nu] = {\rm diag} \{ (-1)^{k_1},~ (-1)^{k_2},~ (-1)^{k_3} \}\, ,
\end{eqnarray}
where $k_{1,2,3}=0,1$.
This conclusion leaves only ${Z}_2$ or ${Z}_2 \times {Z}'_2$ allowed as residual symmetry in the neutrino sector for Majorana neutrinos. 

Together, Eqs.~\eqref{eq:residual_l} and \eqref{eq:residual_nu} lead to the partial dependence of $U_\ell$ and $U_\nu$ on $\rho_{ L} (g_\ell)$ and $\rho_{ L} (g_\nu)$, and eventually result in restriction on the leptonic mixing matrix $U_{\rm PMNS}=U^\dagger_{\ell}U_\nu$. It is convenient to choose a flavor basis that $\rho_{ L}(g_\ell)$ is diagonal, then $U_\ell$ is also diagonal and $U_{\rm PMNS}$ appears to be the unitary matrix to diagonalize $\rho_{ L}(g_\nu)$,  
\begin{align} 
U_{\rm PMNS}^\dag \, \rho_{ L}(g_\nu) \, U_{\rm PMNS} = {\rm diag} \{ (-1)^{k_1},~ (-1)^{k_2},~ (-1)^{k_3} \}.
\end{align}

With the help of residual symmetries, we further show in detail the realization of TM$_1$ mixing pattern from residual symmetries of $S_4$. 
$S_4$ is the permutation group of four objects, which is also isomorphic to the octahedral group. It has 24 elements, generated by three generators satisfying~\cite{King:2013eh}
\begin{align}
    S^2 = T^3 = U^2 = (ST)^3 = (SU)^2 = (TU)^2 =(STU)^4=I \, ,
    \label{eq:s4gen}
\end{align}
where $T$ is an order-3 generator, and $S$ and $U$ are order-2 generators. The group has two irreducible 3-dimensional representations ${\bf 3}$ and ${\bf 3}'$, in which $T$, $S$ and $U$ are represented as
\begin{align} 
\rho_{\mathbf{3}^{(\prime)}}(T) &= \left(
\begin{array}{ccc}
 1 & 0 & 0 \\
 0 & \omega ^2 & 0 \\
 0 & 0 & \omega  \\
\end{array}
\right) , \;  \nonumber \\
\rho_{\mathbf{3}^{(\prime)}}(S) &= \frac{1}{3} \left(
\begin{array}{ccc}
 -1 & 2 & 2 \\
 2 & -1 & 2 \\
 2 & 2 & -1 \\
\end{array}
\right), \; \nonumber \\
\rho_{\mathbf{3},\mathbf{3}'}(U) &= \pm \left(
\begin{array}{ccc}
 1 & 0 & 0 \\
 0 & 0 & 1 \\
 0 & 1 & 0 \\
\end{array}
\right)\, ,
\end{align}
where a basis of diagonal $T$ has been chosen with $\omega=e^{{2\pi i}/{3}}$~\cite{King:2019vhv}. The symmetry ${Z}_3^T = \{ I, T, T^2 \}$ is chosen as the residual symmetry in the charged lepton sector to guarantee a diagonal $U_\ell$. Then, the mixing matrix is only determined by the residual symmetry $G_\nu$ in the neutrino sector. There could be a few choices on the possible $G_\nu$, such as:
\begin{itemize}
\item Taking $G_\nu$ to be $Z_2^{S} = \{I, S \}$, diagonalizing $\rho_{\mathbf{3}^{(\prime)}} (S)$ fixes the second row of $|U_{\rm TBM}|$, i.e., ${\rm TM}_2$ mixing. The first and third row cannot be fixed since both of them refers to the degenerate eigenvalue $-1$ of $\rho_{\mathbf{3}^{(\prime)}} (S)$. Realization of the pattern requires only an $A_4$ symmetry since the generator $U$ is not necessary. Realizations in modular $A_4$ symmetries are presented in \cite{deMedeirosVarzielas:2021pug,Zhang:2024rwv}.

\item Taking $Z_2^{U} = \{I, U \}$, diagonalizing $\rho_{\mathbf{3}^{(\prime)}} (U)$ fixes the third row of $|U_{\rm TBM}|$, leading to $\theta_{13} = 0$ and $\theta_{23}=45^\circ$. This case has already been excluded. 

\item The third choice is to take $G_\nu$ to be $Z_2^{SU} = \{I, SU \}$. Here representation matrices of $SU$ in $\mathbf{3}$ and $\mathbf{3}'$ are explicitly written as 
\begin{eqnarray} 
\rho_{\mathbf{3}, \mathbf{3}'}(SU) = \pm \frac{1}{3} \left(
\begin{array}{ccc}
 -1 & 2 & 2 \\
 2 & 2 & -1 \\
 2 & -1 & 2 \\
\end{array}
\right).
\end{eqnarray}
The vector $(2, -1, -1)$ is the eigenvector respecting the non-degenerate eigenvalue of $\rho_{\mathbf{3}, \mathbf{3}'}(SU)$. 
Therefore, ${\rm TM}_1$ is achieved once the residual symmetry is $Z_2^{SU}$. 
\end{itemize}
The prediction of ${\rm TM}_1$ mixing pattern is thus obtained via the structure of residual symmetries assigned as $G_\ell = Z_3^T$ and $G_\nu = Z_2^{SU}$, which is independent of model details and regardless of realizations in traditional flavor symmetry or modular symmetry. 



\section{Models based on modular $S_4$ symmetry}\label{sec:3}
In this paper, we study phenomenological modular $S_4$ constructions for the lepton sector at a very high energy scale that realize trimaximal mixing through different residual-symmetry breaking patterns. Depending on the setup, the residual symmetries are implemented either by assuming flavon vacuum alignments or by choosing modular fields at fixed points. We do not attempt to derive the full vacuum dynamics or a UV-complete origin of these constructions. We will first review the necessary ingredients for model building based on the modular $S_4$ symmetry, and then follow the approaches given in Figure~\ref{fig:approaches} to construct explicit models featuring TM$_1$ mixing.

Let us begin with the definition of modular $S_4$ symmetry following~\cite{deMedeirosVarzielas:2019cyj,King:2019vhv,Ding:2019gof}. A modular group $\overline{\Gamma}$ is the group of linear fraction transformations acting on the complex modulus $\tau$ in the upper half complex plane (${\rm Im}(\tau)>0$):
\begin{eqnarray}
    \tau \to \gamma \tau = \frac{a\tau+b}{c\tau+d} \, ,
\end{eqnarray}
with $a,b,c,d \in  \mathbb{Z}$ and $ad-bc=1$. When represented by a two by two matrix, $\overline{\Gamma}$ can be expressed as
\begin{eqnarray}
    \overline{\Gamma} = \left\{ \begin{pmatrix}
        a & b \\c &d
    \end{pmatrix}/(\pm {\bf 1})\, , \, a,b,c,d \in  \mathbb{Z} \, , \, ad-bc=1 \right\} \, .
\end{eqnarray}
The modular group is isomorphic to the projective special linear group $PSL(2,\mathbb{Z})=SL(2,\mathbb{Z})/{Z}_2$, and possesses two generators $S_\tau: \tau\to -1/\tau$ and $T_\tau: \tau \to \tau+1$ satisfying $S_\tau^2 = (S_\tau T_\tau)^3={\bf 1}$. By requiring $a,d = 1 ( {\rm mod}  \ N) $ and $b,c = 0 ( {\rm mod}  \ N) $ with $N=2,3,4,\dots$, we can get a subset as
\begin{align}
     \overline{\Gamma} (N)= \left\{ \begin{pmatrix}
        a & b \\c &d
    \end{pmatrix}/ \in PSL(2,\mathbb{Z}) \, , \, \begin{pmatrix}
        a & b \\c &d
    \end{pmatrix} =\begin{pmatrix}
        1 & 0 \\ 0 &1
    \end{pmatrix} ( {\rm mod}  \ N) \right\} \, .
\end{align}

The quotient group $\overline{\Gamma}/\overline{\Gamma}(N)$ can then be defined as the finite modular group $\Gamma_N$, which can be obtained by imposing a constraint $T^N_{\tau}=1$ in addition to $S_\tau^2 = (S_\tau T_\tau)^3={\bf 1}$. The finite modular group $\Gamma_N$ with the first few levels is isomorphic to a permutation group. In particular, it was shown that $\Gamma_2\simeq S_3$, $\Gamma_3\simeq A_4$, $\Gamma_4\simeq S_4$ for $N=2,3,4$~\cite{deAdelhartToorop:2011re}.

For our purpose, we only care about the modular group $S_4$ satisfying $S_\tau^2 = (S_\tau T_\tau)^3=T_\tau^4={\bf 1}$. In literature it is more common to use Eq.~(\ref{eq:s4gen}) to generate the group $S_4$, which is related to the generators of the finite modular group $\Gamma(N)$ as:
\begin{eqnarray}
   S=T_\tau^2\, , \quad  T=S_\tau T_\tau  \, , \quad U=T_\tau S_\tau T_\tau^2 S_\tau \, .
\end{eqnarray}
The CG coefficient of the $S_4$ group is listed in the Appendix of~\cite{Bazzocchi:2009pv}.

In the upper complex plane with the requirement $\tau =\tau+4$, the generators $S$, $T$, and $U$ can be represented by $2\times2$ matrices which are not unique due to the identification. For later convenience, we only list the representations of $T$ and $SU$ here~\cite{deMedeirosVarzielas:2019cyj,King:2019vhv}
\begin{eqnarray}
     T = \begin{pmatrix}
        0 & 1\\-1 & -1
    \end{pmatrix} \, , \quad   SU = \begin{pmatrix}
        -1 & -1\\ 2 & 1
    \end{pmatrix}  \, .
    \label{eq:STUrep}
\end{eqnarray}

If a modulus $\tau_0$ is invariant under a nontrivial $SL(2,Z)$ transformation $\gamma_0\neq I$, then this modulus $\tau_0$ is known as a fixed point which gives $\gamma \tau_\gamma =\tau_\gamma$, with $\gamma$ known as a stabilizer of $\tau_\gamma$. For the $\gamma$ generated by the representations given in Eq.~(\ref{eq:STUrep}), it is straightforward to compute the fixed points:
\begin{eqnarray}
     & 
    \tau_T = \omega=-\frac12+i\frac{\sqrt{3}}{2}\,  , \, \tau_{SU} = -\frac12+\frac{i}{2}\, ,
\end{eqnarray}

Under the modular invariance $\Gamma_N$, a chiral superfield $\phi_i(\tau)$ transforms non-linearly as function of $\tau$ as
\begin{eqnarray}
    \phi_i(\tau) \to \phi_i(\gamma \tau) = (c\tau +d )^{-2k_i} \rho_{I_i}(\gamma) \phi_i(\tau)\, .
\end{eqnarray}
Due to the holomorphicity of the superpotential, the Yukawa coupling transforms as:
\begin{eqnarray}
    Y_{I_Y}(\tau) \to Y_{I_Y}(\gamma \tau)=(c\tau+d)^{2k_Y}\rho_{I_Y}(\gamma) Y_{I_Y}(\tau)  \, ,
\end{eqnarray}
where $k_Y$ must be non-negative, with the weight-zero case corresponding to constant singlet couplings.

If the modular field $\tau$ acquires a VEV at a fixed point such that $\langle \tau \rangle = \tau_\gamma$, an Abelian residual symmetry generated by $\gamma$ is preserved, leaving $Y_I (\gamma\tau_\gamma) = Y_I(\tau_\gamma) $, and thus a characteristic equation can be written
\begin{eqnarray}
    \rho_I(\gamma) Y_I (\tau_\gamma) = (c\tau+d)^{-2k} Y_I (\tau _\gamma) \, ,
\end{eqnarray}
with the eigenvalues $(c\tau+d)^{-2k}$ for the representation matrix $\rho_I (\gamma)$~\cite{Ding:2019gof,deMedeirosVarzielas:2020kji}. Now, given a representation matrix of $S_4$, it is straightforward to use the above equation to determine the eigenvectors for the modular form $Y_I(\tau_\gamma)$. For example, at the fixed point $\tau_\gamma = \tau_T$, the modular forms are~\cite{deMedeirosVarzielas:2019cyj,King:2019vhv}:
\begin{eqnarray}
    & Y_{\bf 3^{(\prime)}}^{(6j+2)}(\tau_T) \propto \begin{pmatrix} 0 \\ 1 \\ 0  \end{pmatrix} \, , \quad   Y_{\bf 3^{(\prime)}}^{(6j+4)}(\tau_T)\propto \begin{pmatrix} 0 \\ 0 \\ 1  \end{pmatrix} \, , \nonumber \\
    & Y_{\bf 3^{(\prime)}}^{(6j+6)}(\tau_T)\propto \begin{pmatrix} 1 \\ 0 \\ 0  \end{pmatrix} \, ,
\end{eqnarray}
where $j$ is a non-negative integer.

Similarly, at the fixed point $\tau_\gamma = \tau_{SU}$, one can solve for the eigenvector of $\rho_{\bf 3'}(SU)$ with respect to the degenerate eigenvalue 1, to obtain the modular forms with weight $\leq 4$ as
\begin{align}
     & Y_{\bf 2}^{(2)}(\tau_{SU})\propto \begin{pmatrix} 1\\ -1 \end{pmatrix} \, , \,  Y_{\bf 3'}^{(2)}(\tau_{SU})\propto \begin{pmatrix} 1\\ 1-\sqrt6 \\ 1+\sqrt6  \end{pmatrix} \, , \nonumber \\ 
     & Y_{\bf 2}^{(4)}(\tau_{SU})\propto \begin{pmatrix} 1\\ 1  \end{pmatrix} \,, \,  Y_{\bf 3}^{(4)}(\tau_{SU})\propto \begin{pmatrix} \sqrt2\\ \sqrt2-\sqrt3 \\ \sqrt2+\sqrt3  \end{pmatrix} \, .
    \label{eq:mf}
\end{align}

With the above knowledge at hand, we can build concrete models based on modular $S_4$ symmetry. Assuming that at low energy, the charged lepton sector and the neutrino sector have different residual flavor symmetries $G_\ell$ and $G_\nu$, respectively, as was discussed in Section~\ref{sec:2.2}, there are at least three approaches to do model building as shown in Figure~\ref{fig:approaches}: 1) the symmetry of charged lepton sector is broken down to a $Z_3^T$ residual symmetry by a flavon $\varphi_l$, while the residual symmetry in neutrino sector is a $Z_2^{SU}$ guaranteed by the VEV of modular field $\tau_\nu$; 2) the symmetry of charged lepton sector is broken down to a $Z_3^T$ residual symmetry due to the modular field $\tau_l$, while a $Z_2^{SU}$ residual symmetry remains in the neutrino sector due to the VEVs of flavons $\varphi_\nu$ that only couple to neutrinos; 3) the sectors of charged leptons and neutrinos are dependent on two different modular fields, and the final trimaximal mixing pattern is achieved by choosing specific VEVs for the modular fields. In the following subsections, we will proceed with analyzing three different models explicitly with the mentioned approaches, and see how they result in similar neutrino mass matrices and phenomenology.

\subsection{Model A}\label{sec:3.1}
The first model realizing the TM$_1$ mixing from modular $S_4$ symmetry can be achieved by adding flavons in the lepton sector that breaks the flavor symmetry to the residual $Z_3^T$, while adding a modular field $\tau_\nu$ in the neutrino sector whose VEV is chosen at $\langle \tau_\nu \rangle = \tau_{SU}$  to ensure the residual $Z_2^{SU}$. By this setup, all neutrino couplings are promoted to modular forms, making the neutrino sector more economical at the phenomenological level. 

However, the most economical realization of this setup, following directly the schematic construction in Fig.~1(A), is already quite restrictive and requires additional auxiliary ingredients because of the following two reasons.
Firstly, there are only two flavor singlets under $S_4$, either ${\bf 1}$ or ${\bf 1'}$, so it is hard to embed three right-handed charged leptons within the two non-degenerate flavor singlets, unless an auxiliary symmetry like a $Z_n$ is introduced to separate the charged leptons. Though this problem may be avoided by introducing a modular weight to one of the degenerate charged leptons, $Z_n$ charges are still needed if one requires the absence of trivial neutrino mass terms. Secondly, the contraction between a flavon $\varphi_l \sim {\bf 3}$ and left-handed charged leptons $L\sim{\bf 3}$ cannot contract with the singlet ${\bf 1}'$ as the decomposition of representation products gives: ${\bf 3}\times{\bf 3}\to {\bf 1}+{\bf 2}+{\bf 3}+{\bf 3}'$. This means that one needs to introduce an extra flavon, such as $\eta\sim {\bf 2}$, to make them contractible. These couplings will inevitably be higher-order operators as can be checked from the superpotential in Eq.~(\ref{eq:WmodelA}). To cancel all lower-order couplings while keeping all the first three terms in Eq.~(\ref{eq:WmodelA}), there will be a constraint on the charge of the auxiliary discrete symmetry $Z_n$. 

\begin{table}[h]
    \centering
    \renewcommand{\arraystretch}{1.2}
    \begin{tabular}{|c|c c ||c|c c ||c|c c|}
    \hline
        Fermions & $S_4$ &  $2k$ & Scalars & $S_4$ & $2k$ & Modular forms & $S_4$ & $2k$ \\
        \hline
        $e^c$  & {${\bf 1}$}  & $0$  & $\varphi_l$ &  {${\bf 3}$}  & $+2$ &  $Y_{\bf 2}(\tau_\nu)$ &   {${\bf 2}$} &  $+2$ \\
        $\mu^c$ & {${\bf 1'}$}  & $0$ & $\eta$ &  {${\bf 2}$}  & $+2$ & $Y_{\bf 3'}(\tau_\nu)$ &   {${\bf 3'}$} & $+2$   \\
        $\tau^c$ & {${\bf 1}$}  & $0$ & $\xi$ &  {${\bf 1'}$}  & $+6$  & & &  \\
        $L$ &  {${\bf 3}$} & $0$ &  $H_{u}$ &  {${\bf 1}$} & $+4$ & & &  \\ 
        $\nu^c$ & {${\bf 3}$}  & $-4$  &  $H_{d}$  & {${\bf 1}$} & $-4$ & & &\\ 
        \hline
    \end{tabular}
    \caption{The $S_4$ representations and the modular weights of different superfields and Yukawa couplings in Model A}
    \label{tab:Lepton and Higgs superfields.}
\end{table}

Now we can write down a simple toy model by assigning the $S_4$ representations and modular weights as shown in Table~I, together with suitable auxiliary $Z_n$ charges, so that the following operators are the leading terms in the superpotential:
\begin{align}
    W =&   \frac{y_e}{\Lambda^2}(L\varphi_l^2)_{\bf 1}e^c H_d + \frac{y_\mu}{\Lambda^2}  (L(\varphi_l\eta)_{\bf 3'})_{\bf 1'} \, \mu^c H_d    \nonumber \\
    &+\frac{y_\tau}{\Lambda^2} (L(\varphi_l \eta)_{\bf 3})_{\bf 1}\, \tau^c H_d +\ y_D  L\nu^cH_u    \nonumber \\
    & + \frac12 Y_{\bf 2}(\tau_\nu) (\nu^c\nu^c)_{\bf 2} \xi +\frac12 Y_{\bf 3'}(\tau_\nu) (\nu^c\nu^c)_{\bf 3} \xi\, ,
    \label{eq:WmodelA}
\end{align}
where $L$ is the left-handed charged lepton doublets, $e^c$, $\mu^c$, $\tau^c$ are right-handed charged leptons, $\nu^c$ are right-handed neutrinos, and a few flavons $\varphi_l$, $\eta$, $\xi$ are introduced to make them contractible. 

The non-renormalizable terms in Eq.~(\ref{eq:WmodelA}) should be understood in an effective-field-theory sense. In particular, they can be generated after integrating out heavy vector-like messenger superfields with masses of order $\Lambda$, in close analogy with Froggatt-Nielsen constructions. A more complete UV realization including these vectorlike fields may require more auxiliary Abelian symmetries to remove those unwanted lower-order operators, but specifying such a unique completion is not the purpose of the present phenomenological study.

At the phenomenological level, we therefore adopt the flavon VEVs $\langle \varphi_l \rangle = v_{T} ( 1,0 ,0 )^T$, $\langle \eta \rangle=  (v_{\eta_1}, v_{\eta_2})^T $, and $\langle \xi \rangle \neq 0$ so as to preserve the residual $Z_3^T$ symmetry in the charged lepton sector, following the standard residual-symmetry approach in the $S_4$ flavor-model literature \cite{Luhn:2013vna,King:2013eh}. Likewise, we choose  $\langle \tau_\nu \rangle =\tau_{SU} = -\frac12+\frac{i}{2}$ to preserve the residual $Z_2^{SU}$ symmetry in the neutrino sector. From a UV-complete perspective, the above alignment can in principle be derived by introducing driving fields with suitable modular weights and constructing the full flavon superpotential, so that the vacuum alignment follows from vanishing driving-field $F$-terms, as in conventional supersymmetric $S_4$ models~\cite{Bazzocchi:2009pv,Ding:2009iy}. Nevertheless, a fully explicit construction that uniquely singles out the required operators may still require a sufficiently large auxiliary $Z_n$ symmetry together with suitable modular-weight assignments (see Appendix \ref{app:vacuum-alignment} as an example). Since this model-building aspect lies beyond the scope of the present work and does not affect the subsequent low-energy analysis once the alignment is taken as an input, we do not attempt a unique complete vacuum construction here, and we will give a possible realization including the flavon sector in Appendix \ref{app:vacuum-alignment} for reference.  

As an illustrative consistency check relevant to Eq.~(\ref{eq:WmodelA}), one possible minimal choice is a $Z_5$ symmetry with $Z_5(L)=Z_5(\varphi_l)=1$, $Z_5(e^c)=Z_5(\xi)=2$, $Z_5(\eta)=3$, and $Z_5(\nu^c)=4$, while all other fields are neutral. This assignment keeps the three $1/\Lambda^2$ charged-lepton operators as the leading terms and forbids all other lower-order and same-order competing terms.

In the charged lepton sector, the mass matrix includes a $\mu-\tau$ mixing as
\begin{eqnarray}
    M _{\ell}^* = \frac{v_d v_{T} }{\Lambda^2}\begin{pmatrix}  y_ev_{T} & 0 & 0 \\ 0 & y_\mu v_{\eta_1} & y_\tau v_{\eta_1} \\ 0 & -y_\mu v_{\eta_2} & y_\tau v_{\eta 2}  \\
    \end{pmatrix} \, .
\end{eqnarray}
where $v_d=\langle H_d \rangle$, and the charged lepton mass matrix is defined in the left-right convention to be consistent with the convention used in Eq.~(\ref{eq:Lleptons}), which gives an additional complex conjugation~\cite{King:2021fhl}. Thus, a unitary matrix $U_\ell$ 
\begin{align}
    U_\ell =& \begin{pmatrix}
        1 & 0 & 0 \\ 0 & \cos \theta_\ell \,e^{i\phi_\ell}  & \sin \theta_\ell\\ 
        0 & -\sin \theta_\ell & \cos \theta_\ell \, e^{-i\phi_\ell}
        \label{eq:UL}
        \end{pmatrix}  \nonumber \\
        \equiv& \begin{pmatrix} 
        e^{-i\phi_3'} & 0 & 0 \\ 0 & \cos \theta_\ell \,e^{i\phi_1}  & \sin \theta_\ell \, e^{-i\phi_2} \\
        0 & -\sin \theta_\ell \, e^{i\phi_2} & \cos \theta_\ell \, e^{-i\phi_1}
        \end{pmatrix} \nonumber \\
        &\times\begin{pmatrix}
        e^{i\phi_3'} & 0 & 0 \\ 0 & e^{-i\phi_2}  & 0 \\ 
        0 & 0 &  e^{i\phi_2} 
        \end{pmatrix}
\end{align}
 should be included to diagonalize $M_\ell M_\ell^\dag$, where $\phi_3'$ and $\phi_2$ can be absorbed by re-defining the right-handed charged leptons.

In the neutrino sector, the Dirac mass matrix in the flavor basis, i.e., the basis of charge lepton mass eigenstates, is given simply by
\begin{eqnarray}
    M_D = y_D^* v_u U_\ell^\dagger \begin{pmatrix}
        1 & 0 & 0 \\ 0 & 0 & 1 \\ 0 & 1 & 0
    \end{pmatrix} \, ,
    \label{eq:MD}
\end{eqnarray}
where $v_u \equiv \langle H_u \rangle  $, and $U_\ell^\dagger$ is inserted to diagonalize the charged leptons.

Considering the Majorana neutrino mass matrix, we start with the simplest case by including only weight-2 modular forms, $Y_{\bf 2}(\tau_\nu)\sim{\bf 2}$ and $Y_{\bf 3'} (\tau_\nu)\sim{\bf 3'}$. Apparently, making the modular weights higher than 2 will increase the number of free parameters that will fit with the data eventually, which is the reason why we only check how the model can be built with the least number of free parameters. Using Eq.~(\ref{eq:mf}) the value of $Y_{{\bf 2}}^{(2)}(\tau_{SU})$ and $Y_{{\bf 3'}}^{(2)}(\tau_{SU})$ for the weight $2k_\nu=2$, the Majorana mass matrix is
\begin{align}
   \! M_R^* = \! \begin{pmatrix}
        0 & -Y_{{\bf 2},1} & Y_{{\bf 2},2} \\ -Y_{{\bf 2},1} & Y_{{\bf 2},2} & 0 \\ Y_{{\bf 2},2} & 0 & -Y_{{\bf 2},1}   \end{pmatrix} \!
        +\! \begin{pmatrix}
        2Y_{{\bf 3'}\! ,1} & -Y_{{\bf 3'}\! ,3} & -Y_{{\bf 3'}\! ,2} \\ -Y_{{\bf 3'}\! ,3} & 2Y_{{\bf 3}'\! , 2} & -Y_{{\bf 3'} \! ,1} \\ -Y_{{\bf 3'}\! , 2} & -Y_{{\bf 3'}\! ,1} & 2Y_{{\bf 3'}\! , 3} \end{pmatrix} \! .
\end{align}

It is more convenient to re-parameterize the above Majorana mass matrix into the following form:
\begin{eqnarray}
    M_R = & 
    a_2\begin{pmatrix}
         0 & 1 & 1 \\ 1 & 1 & 0 \\ 1 & 0 & 1 \end{pmatrix} 
     + a_3\begin{pmatrix}
         2 & -1 & -1 \\ -1 & 2 & -1 \\ -1 & -1 & 2 \end{pmatrix} \nonumber \\
    &- a_3\sqrt{6}\begin{pmatrix}
         0 & 1 & -1 \\ 1 & 2 & 0 \\ -1 & 0 & -2 \end{pmatrix} \, ,
    \label{eq:Mass-Maj1}
\end{eqnarray}
where $a_2 =[\lambda_2 Y^{(2)}_{{\bf 2},1}(\tau_{SU})]^* , a_3 = [\lambda_3 Y^{(2)}_{{\bf 3'},1}(\tau_{SU})]^*$ are complex coefficients. 

Finally, by applying the standard neutrino seesaw formula
\begin{eqnarray}
    M_\nu = - M_D M_R^{-1} M_D^T  \, ,
    \label{eq:seesaw}
\end{eqnarray}
it is straightforward to compute the mass matrix for the active neutrinos and then proceed with the diagonalization to find the PMNS matrix, which will be discussed in details in the next section.

\subsection{Model B}\label{sec:3.2}
An alternative approach to the previous scenario is to use the VEV of the modular field to achieve $Z_3^T$ in the charged lepton sector, and then introduce additional flavons in the neutrino sector to get the residual $Z_2^{SU}$ symmetry. This scenario should be similar to the method discussed in~\cite{Luhn:2013lkn}. The superfield assignments of Model B are listed in the Table~\ref{tab:2}.
\begin{table}[h]
    \centering
    \renewcommand{\arraystretch}{1.2}
    \begin{tabular}{|c|c c||c|c c||c|cc|}
    \hline
        Fermions & $S_4$ &  $2k$ & Scalars & $S_4$ &  $2k$ & Modular forms & $S_4$ & $2k$  \\
        \hline
        $e^c$  & {${\bf 1'}$}  & $-4$ & $\varphi_\nu$ &  {${\bf 3}$} & $+4$ & $Y_e(\tau_l)$  &  {${\bf 3'}$} & $+6$ \\
        $\mu^c$ & {${\bf 1'}$}  & $-2$ & $\varphi_\nu'$ &  {${\bf 3'}$} & $+3$ & $Y_\mu(\tau_l)$  &  {${\bf 3'}$} & $+4$ \\
        $\tau^c$ & {${\bf 1'}$} & $0$ &  $\eta$ & {${\bf 1'}$} &  $+1$ & $Y_\tau(\tau_l)$  &  {${\bf 3'}$} & $+2$ \\
        $L$ &  {${\bf 3}$} & $0$  &  $\xi$ &  ${\bf 1'}$ & $+2$  &   &   &   \\ 
        $\nu^c$ & {${\bf 3}$}  & $-2$ & $H_{u}$ & {${\bf 1}$} & $+2$ & & &  \\
          &    &   &  $H_{d}$ &  {${\bf 1}$} & $-2$  &  &  &   \\ 
        \hline
    \end{tabular}
    \caption{The superfields and Yukawa couplings with different representations and weights in Model B. }
    \label{tab:2}
\end{table}

The superpotential that is invariant under this superfield assignment is:
\begin{align}
    W =&   Y_e(\tau_l) Le^c H_d+Y_\mu(\tau_l) L\mu^c H_d +Y_\tau(\tau_l) L\tau^c H_d \nonumber \\
    & + \ y_{D} (L\nu^c)_{\bf1}H_u + \frac{y_M}{2\Lambda} 
    \xi^2 (\nu^c\nu^c)_{\bf 1} \nonumber \\
    & + \frac{\lambda}{2} \varphi_{\nu}  (\nu^c\nu^c)_{\bf 3} + \frac{\lambda'}{2\Lambda} \varphi_{\nu}'  (\nu^c\nu^c\eta)_{\bf 3'}  \, .
    \label{eq:superpotential_B}
\end{align}

As in Model A, the higher-dimensional terms are interpreted within a flavor EFT with cutoff $\Lambda$. To forbid the leading Majorana mass terms like $M_{\bf 1}(\nu^c\nu^c)$, $Y_{\bf 3'}(\nu^c\nu^c)_{\bf 3}\xi$ and $Y_{\bf 2}(\nu^c\nu^c)_{\bf 3}\varphi_\nu'$, an auxiliary $Z_n$ symmetry is required with non-trivially charged neutrinos. In Appendix~\ref{app:vacuum-alignment}, we give an explicit realization of the superpotential from the flavon driving fields using a minimal $Z_5$ symmetry.

When the VEV of modular field is set to the fixed point $\tau_T$, the charged leptons will be put into the AF basis~\cite{King:2019vhv}, implying
\begin{align}
    Y_e(\tau_T) \propto \begin{pmatrix} 1 \\ 0 \\ 0 \end{pmatrix}  , \,  Y_\mu(\tau_T) \propto \begin{pmatrix} 0 \\ 0 \\ 1\end{pmatrix} , \,  Y_\tau(\tau_T) \propto \begin{pmatrix} 0 \\ 1 \\ 0 \end{pmatrix} ,
\end{align}
so that the lepton mass matrix is diagonal. 

In the neutrino sector, the Dirac neutrino mass matrix has the same form as in Eq.~(\ref{eq:MD}). In the Majorana terms, the right-handed neutrino triplets $(\nu^c \nu^c)$ in this model should be contracted following the $S_4$ product rule ${\bf 3}\times {\bf 3} = {\bf 1}+{\bf 2}+{\bf 3}+{\bf 3'}$. However, as shown in~\cite{Luhn:2013lkn}, if we have only one flavon field which is a triplet ${\bf 3}$ of $S_4$, the deduced Majorana mass matrix leads to only a tri-bimaximal mixing with $\theta_{13}=0$ predicted. Therefore, an economical choice is to introduce only two flavons in either a ${\bf 3}$ or a ${\bf 3'}$ representation of $S_4$. Following the standard residual-symmetry model-building strategy, we adopt the flavon alignments
\begin{eqnarray}
   \langle \varphi_\nu \rangle = v_{SU}\begin{pmatrix}   1 \\ 1 \\ 1 \end{pmatrix} \, , \,  \langle \varphi_\nu ' \rangle = v_{SU}'\begin{pmatrix}   0 \\ 1 \\ -1 \end{pmatrix} \, .
\end{eqnarray}
which preserve the residual symmetry $Z_2^{SU}$. A possible realization of such vacuum alignment, including the flavon sector consistent with the auxiliary $Z_n$ charge assignment that guarantees the operators in eq.~(\ref{eq:superpotential_B}) are the leading ones, is provided in detail in Appendix \ref{app:vacuum-alignment}.

Expanding the superpotential, the Majorana mass matrix for the right-handed neutrinos now reads
\begin{eqnarray}
    M_R =& b_1\begin{pmatrix} 
        1 & 0 & 0 \\ 0 & 0 & 1 \\ 0 & 1 & 0 \end{pmatrix} + b_2 \begin{pmatrix}
             2 & -1 & -1 \\ -1 & 2 & -1 \\ -1 & -1 & 2
        \end{pmatrix} \nonumber \\
    & +  b_3 \begin{pmatrix}
         0 & 1 & -1 \\ 1 & 2 & 0 \\ -1 & 0 & -2
    \end{pmatrix} \, ,
    \label{eq:Mass-Maj2}
\end{eqnarray}
with $b_1 = \left(\frac{y_M v_\xi^2}{\Lambda}\right)^*$, $b_2 = (\lambda v_{SU})^*$ and $b_3 =\left( \frac{\lambda' v_{SU}'v_\eta}{\Lambda}\right)^*$. Due to the suppression of higher-dimensional operator, $b_3$ is in general much smaller than $b_2$. Similar to Model A, the complex conjugate appears because of a matching to the $M_R$ defined in the left-right convention. To this point, all the mass matrices have been written so the seesaw equation Eq.~(\ref{eq:seesaw}) can be applied again to find the mass matrix for the active neutrinos.


\subsection{Model C}\label{sec:3.3}
The previous two scenarios contain only one modular field in either the lepton or the neutrino sector, and a few flavons with particular VEVs alignment to obtain the desired residual symmetry. It is also possible to use two modular fields, $\tau_l$ and $\tau_\nu$, in both the lepton and the neutrino sector, so that all Yukawa couplings can be expressed as modular forms. The generalization from one modulus to multiple moduli was given in~\cite{deMedeirosVarzielas:2019cyj,King:2019vhv}, where a bi-triplet flavon $\Phi$ was introduced to break the modular $S_4^l \times S_4^\nu$ symmetry down to a diagonal $S_4$ subgroup through the VEV of $\Phi$. The two modular fields can get different VEVs at different fixed points which then result in different residual symmetries in the lepton or neutrino sectors. A minimal model realizing the above picture requires the superfields to be assigned according to Table~\ref{tab:superfields-scenario-3} ~\cite{King:2019vhv}.  

\begin{table}[h]
    \centering
    \renewcommand{\arraystretch}{1.2}
    \begin{tabular}{|c|c c c c||c|c c c c|}
    \hline
        Fields & $S_4^l$ & $S_4^\nu$ & $2k_l$ & $2k_\nu$ & \makecell{Yukawas/\\ masses} & $S_4^l$ & $S_4^\nu$ & $2k_l$ & $2k_\nu$ \\
        \hline
        $e^c$  & {${\bf 1'}$} & {${\bf 1}$} & $-6$ & $-2$ & $Y_e(\tau_l)$  & {${\bf 3'}$} & {${\bf 1}$} & $+6$ & 0 \\
        $\mu^c$ & {${\bf 1'}$} & {${\bf 1}$} & $-4$ & $-2$ & $Y_\mu(\tau_l)$ & {${\bf 3'}$} & {${\bf 1}$} & $+4$ & 0 \\
        $\tau^c$ & {${\bf 1'}$} & {${\bf 1}$} & $-2$ & $-2$ & $Y_\tau(\tau_l)$ & {${\bf 3'}$} & {${\bf 1}$} & $+2$ & 0 \\
        \cline{6-10}
        $L$ & {${\bf 3}$} & {${\bf 1}$} & 0 & $+2$ & $M_1(\tau_\nu)$ & {${\bf 1}$} & {${\bf 1}$} & 0 & $+4$ \\
        $\nu^c$ &  {${\bf 1}$} & {${\bf 3}$} & 0 & $-2$ & $M_2(\tau_\nu)$ &  {${\bf 1}$} & {${\bf 2}$} & 0 & $+4$ \\
        \cline{1-5}
        $\Phi$ &  {${\bf 3}$} & {${\bf 3}$} & 0 & 0 & $M_3(\tau_\nu)$ &  {${\bf 1}$} & {${\bf 3}$} & 0 & $+4$ \\
        $H_{u,d}$ &  {${\bf 1}$} & {${\bf 1}$} & 0 & 0 & & & & & \\
        \hline
    \end{tabular}
    \caption{Representations and modular weights for leptons, flavon, Higgs fields, Yukawa couplings and Majorana masses in Model C.}
    \label{tab:superfields-scenario-3}
\end{table}

The superpotential for these superfields is given by:
\begin{align}
    W &=   Y_e(\tau_l) Le^c H_d+Y_\mu(\tau_l) L\mu^c H_d +Y_\tau(\tau_l) L\tau^c H_d  \nonumber \\
    & + \frac{y_\nu}{\Lambda} L\Phi H_u \nu^c +\frac12 M_{\bf 1}(\tau_\nu) (\nu^c\nu^c)_{\bf 1}  \nonumber \\
    & +\frac12 M_{\bf 2}(\tau_\nu) (\nu^c\nu^c)_{\bf 2}+\frac12 M_{\bf 3}(\tau_\nu) (\nu^c\nu^c)_{\bf 3}\, .
\end{align}
In our paper, the two modular fields $\tau_l$ and $\tau_\nu$ are taken as independent inputs whose VEVs are chosen at fixed points associated with the residual symmetries in the charged-lepton and neutrino sectors. Multiple-moduli frameworks and explicit stabilization mechanisms have been discussed in the literature \cite{Kobayashi:2019xvz,deMedeirosVarzielas:2019cyj,King:2019vhv,Novichkov:2022wvg,King:2023snq}, but deriving the vacuum structure and any relation between $\tau_l$ and $\tau_\nu$ from compactification of extra dimensions is outside the scope of the present
phenomenological analysis.
This scenario gives the same Yukawa matrices for charged leptons when the modular field $\tau_l$ is chosen at $\langle \tau_l \rangle = \tau_T$ and thus will not be repeated.

Following the construction of Ref. \cite{King:2019vhv}, the symmetry breaking $S_4^l \times S_4^\nu \to S_4$ is naturally achieved via the VEV of a bi-triplet scalar field $\Phi\sim ({\bf 3}, {\bf 3})$ of $S_4^l \times S^\nu_4$. By minimizing the superpotential, the scalar gains a VEV along the direction of $\langle \Phi \rangle _{\alpha i}=v_{\Phi} (P_{23})_{\alpha i}$ with
\begin{eqnarray}
    P_{23} = \begin{pmatrix}
        1 & 0 & 0 \\ 0 & 0 & 1 \\ 0 & 1 & 0 
    \end{pmatrix} \, . 
\end{eqnarray}
Expanding the Dirac Yukawa coupling $ \frac{y_\nu}{\Lambda} L\Phi H_u \nu^c$ after the breaking of $S_4^l \times S_4^\nu \to S_4$ and the Higgs $H_u$ acquiring a VEV, we get
\begin{eqnarray}
    M_D =  y_D^* P_{23} v_u \, , \quad {\rm with} \quad y_D = \frac{y_\nu v_\Phi}{\Lambda} \, .
\end{eqnarray}

The Majorana neutrino sector includes three modulus-dependent couplings $M_{\bf r}(\tau_\nu)$ with ${\bf r} = {\bf 1,2,3}$ that in general give a Majorana mass matrix with the following form~\footnote{Note that there are no $M_{\bf 3'}$ term since it vanishes because of its antisymmetric nature.}:
\begin{eqnarray}
        M_R^* =& \begin{pmatrix}
        M_{\bf 1} & 0 & 0 \\ 0 & 0 & M_{\bf 1} \\ 0 & M_{\bf 1} & 0   \end{pmatrix} + \begin{pmatrix}
        0 & M_{{\bf 2},1} & M_{{\bf 2},2} \\ M_{{\bf 2},1} & M_{{\bf 2},2} & 0 \\ M_{{\bf 2},2} & 0 & M_{{\bf 2},1}   \end{pmatrix}   \nonumber \\
        &+\begin{pmatrix}
        2M_{{\bf 3},1} & -M_{{\bf 3},3} & -M_{{\bf 3},2} \\ -M_{{\bf 3},3} & 2M_{{\bf 3},2} & -M_{{\bf 3},1} \\ -M_{{\bf 3},2} & -M_{{\bf 3},1} & 2M_{{\bf 3},3}   \end{pmatrix} \, .
        \label{eq:massRN} 
\end{eqnarray}
Assuming that the modular field $\tau_\nu$ is fixed at the fixed point $\langle \tau_\nu \rangle= \tau_{SU}$, we can use the modular forms given in Eq.~(\ref{eq:mf}) to rewrite  the Majorana mass matrix $M_R$, and then parameterize it into the following form~\cite{King:2019vhv}:
\begin{align}
    &M_R  =  c_1\begin{pmatrix} 
        1 & 0 & 0 \\ 0 & 0 & 1 \\ 0 & 1 & 0 \end{pmatrix} \!+ 
        c_2\begin{pmatrix}
             0 & 1 & 1 \\ 1 & 1 & 0 \\ 1 & 0 & 1 \end{pmatrix}  \nonumber \\
         &+ c_3 \sqrt{2}\begin{pmatrix}
             2 & -1 & -1 \\ -1 & 2 & -1 \\ -1 & -1 & 2 \end{pmatrix} \!
        - c_3 \sqrt{3}\begin{pmatrix}
         0 & 1 & -1 \\ 1 & 2 & 0 \\ -1 & 0 & -2 \end{pmatrix} \! \, ,
    \label{eq:Mass-Maj3}
\end{align}
where $c_1 = [M_{\bf 1} (\tau_{SU})]^*$, $c_2 = [M_{{\bf 2},1} (\tau_{SU})]^*$ and $c_3 =\frac{1}{\sqrt2} [M_{{\bf 3},1} (\tau_{SU})]^*$. 

Thus the active neutrino mass matrix becomes
\begin{eqnarray}
    M_\nu = - M_D M_R^{-1} M_D^T = -(y_D^*)^2 v_u^2 P_{23} M_R^{-1} P_{23} \, .
\end{eqnarray}

\section{Fit with JUNO Data}\label{sec:4}
\subsection{Parameterization}
With the lepton and neutrino mass matrices given in the above models, it is now straightforward to compute the experimental predictions for Majorana mass matrices and mixings. In particular, since all three models exhibit a similarity in their mass matrices, it is convenient to first put them into a block diagonal form by applying a TBM mixing matrix,
\begin{eqnarray}
     U_{\rm TBM}^T M_R U_{\rm TBM} = 
     \begin{pmatrix}
        \eta & 0 & 0 \\ 0 & \alpha & \gamma \\ 0 & \gamma & \beta 
    \end{pmatrix} \, ,
    \label{eq:parameterization}
\end{eqnarray}
where it can be shown that
\begin{align}
        {\rm Model \ \ A:} & \quad  \eta= -\alpha + \beta\,,  ~ \gamma = \alpha - 2\beta, \nonumber \\ 
        &  \text{with}\quad  
        \alpha = 2 a_2, ~\beta = a_2 + 3 a_3\, ,  \\ 
        {\rm Model \ \ B:} &  \quad \eta = 2 \alpha + \beta, \nonumber \\ 
        & \hspace{-4.5mm} \text{with}\quad  
        \alpha= b_1, ~ \beta = -b_1 + 3b_2 \, ,~ \gamma= \sqrt6 b_3 \, ,  \\ 
        {\rm Model \ \ C:} & \quad \eta = -\beta - 2 \gamma, \nonumber \\ 
         & \hspace{-4.5mm} \text{with}   \quad
         \alpha = c_1 + 2 c_2, ~ \beta = -c_1 + c_2 + 3 \sqrt2 c_3 , \nonumber \\   
         & \qquad \gamma = -3 \sqrt2 c_3 \, ,
\end{align}
where the parameters $a_i$, $b_i$, and $c_i$ with $i=1,2,3$ are free parameters defined for Models A, B, and C in Eqs.~(\ref{eq:Mass-Maj1}), (\ref{eq:Mass-Maj2}), (\ref{eq:Mass-Maj3}), respectively. Therefore, we can always simplify the computation by re-parameterizing those free parameters in the original $M_R$ into the above form with either two complex parameters ($\alpha$, $\beta$ for Model A), or three complex parameters ($\alpha,\,\beta,\,\gamma$ for Models B and C).

On the other hand, the Dirac mass matrix takes a similar form except for a unitary matrix $U_\ell$ particularly for Model A in the charged lepton flavor basis. Using the type-I seesaw formula in eq.~(\ref{eq:seesaw}), we can write
\begin{eqnarray}
    M_\nu = -m_0^2 U_\ell^\dagger P_{23} M_R^{-1} P_{23}^T U_\ell^*\, . 
\end{eqnarray}
with an overall factor $m_0^2 = (y_D^*)^2 v_u^2$. 

The above neutrino mass matrix in the flavor basis can be diagonalized by the PMNS matrix defined as $U_{\rm PMNS}  = U^\dagger_{\ell}U_{{\rm TM}_1}$ with $U_\ell$ and $U_{{\rm TM}_1}$ given in Eq.~(\ref{eq:UL}) and Eq.~(\ref{eq:TM1}) 
\begin{align}
    U_{{\rm PMNS}} ^\dagger M_\nu  U_{{\rm PMNS}}^*  =& -m_0^2 \nonumber \\
    \times
    &\begin{pmatrix}
    \frac{1}{|\eta|} & 0 \\ 0 & e^{-2i \alpha_3} \left[V^\dagger \begin{pmatrix}
        \alpha & \gamma \\ \gamma & \beta
    \end{pmatrix} V^*\right]^{-1}     
    \end{pmatrix} 
    \nonumber \\
    \equiv&  -m_0^2
    \begin{pmatrix}
        M_1^{-1} & & \\ & M_2^{-1} & \\ & & M_3^{-1}
    \end{pmatrix} \!,
    \label{eq:diag-mnu}
\end{align}
where a phase factor of $\eta$ can be extracted to reduce one free parameter, and the unitary matrix $V$ is used to diagonalize the bottom-right block defined as
\begin{eqnarray}
    V \equiv e^{i\alpha_3} \begin{pmatrix}
         \cos\theta_R e^{-i\alpha_1} & \sin \theta_R e^{i\alpha_2} \\
         \sin \theta_R e^{-i\alpha_2} & -\cos\theta_R e^{i\alpha_1}
         \end{pmatrix} \, .
\end{eqnarray}
Under this parameterization, the general form of $U_{{\rm TM}_1}$ can be expressed in terms of the TBM matrix defined in Eq.~(\ref{eq:TBM}) multiplied by a unitary matrix:
\begin{align}
    U_{{\rm TM}_1} \equiv U_{\rm TBM} 
      \begin{pmatrix}
        e^{-i \alpha_3'} & 0 & 0 \\  
        0 & \cos\theta_R \, e^{i\alpha_1} & \sin \theta_R \, e^{-i\alpha_2} \\
        0 & -\sin \theta_R \, e^{i\alpha_2} & \cos\theta_R\,  e^{-i\alpha_1}
    \end{pmatrix} 
    ,
    \label{eq:UTM}
\end{align}
where $\alpha_3'=\frac12 {\rm arg}(\eta)$. 

As a result, the predictions of observables under the above parameterization, such as the mixing angles and the CP-violating phase, can be computed explicitly. Since an overall phase of $\gamma$ can be dropped, there are at most five free parameters for all three proposed models to fit with the five observables in neutrino experiments including three mixing angles and two mass differences.

\subsection{Predictions of observables and sum rules}
A straightforward result from the above parameterization gives the predictions:
\begin{eqnarray}
    &&\sin \theta_{13} = \frac{\sin \theta_R }{\sqrt{3}} \, , \quad     \label{eq:observables1.1} \\
    &&\tan \theta_{12} = \frac{\cos \theta_R }{\sqrt{2}} \, , \quad     \label{eq:observables1.2} 
\end{eqnarray}
These correlations recover sum rules between mixing angles shown in Eq.~(\ref{eq:sumrule}).

In Model A, the mixing angle $\theta_{23}$ and CP-violating phase $\delta$ will depend on the two physical parameters $\theta_\ell$ and $\phi_\ell$ in the mixing matrix $U_\ell$, so they will not be expanded here. The condition $\gamma=\alpha-2\beta$ reduces the original five parameters in Eq.~(\ref{eq:parameterization}) to three parameters including $|\alpha|$, $|\beta|$ and ${\rm arg}(\beta)$ in $M_R$. After re-parameterizing them by $\theta_R$, $\alpha_1$, $\alpha_2$ and $m_i\equiv m_0^2 M_i^{-1}$ for $i=1,2,3$, two strong correlations were found between the mixing angle $\sin\theta_{13}$ and three eigenvalues of neutrino mass matrix:
\begin{align}
   \sin\left(2\theta_R\right)\cos\left(\alpha_1-\alpha_2\right) 
    = 1 + \frac{\left( 5 m_2^2 - 2m_1^2 \right) m_3^2}{m_1^2 \left( m_3^2 -m_2^2 \right)} \, ,
   \label{eq:sumrule-A}
\end{align}
\begin{align}
 1 =& -\left[\frac{\Delta m_{21}^2}{4m_1^2} 
          - \frac{\Delta m_{32}^2 \left(2 - 5\sin^2 \theta_R\right)}{8m_3^2} + \frac18\right]^2  \nonumber \\
    &+ \frac{m_3^4}{m_1^4}\ \frac{20  \left( \Delta m_{21}^2 \right)^2 + 6 m_1^4}{\left(  \Delta m_{32}^2 \right)^2 \sin^2 \left( 2\theta_R \right) } \nonumber \\
    & + \ \frac{m_3^2}{m_1^2}\left( \frac{22 m_3^2 \Delta m_{21}^2 }{\left( \Delta m_{32}^2 \right)^2 \sin^2 \left( 2\theta_R \right) }\right. 
    \nonumber\\ &
    + \left. \frac{ 4\Delta m_{21}^2 - 2m_2^2 
          \left(3\sin^2 \theta_R - 4\right)}{ \Delta m_{32}^2  \sin^2 \left( 2\theta_R \right) } \right)\,. 
\label{eq:sumrule-A2}
\end{align}

Given the 3$\sigma$ range of $\sin\theta_{13}$, $\Delta m_{21}^2$, and $\Delta m_{31}^2$ from the NuFIT data~\cite{Esteban:2024eli}, the Eqs.~(\ref{eq:observables1.1})-(\ref{eq:sumrule-A2}) can be solved together.
The normal ordering (NO) case is clearly excluded because the right-hand side of Eq.~(\ref{eq:sumrule-A}) gives a value larger than 1. 
For the inverted ordering (IO) case, $m_3^2-m_2^2<0$ so the Eq.~(\ref{eq:sumrule-A}) implies 
 \begin{eqnarray}
     1-\sin\left(2\theta_R\right)\cos\left(\alpha_1-\alpha_2\right) \approx \frac{3m_3^2}{m_2^2-m_3^2}\, ,
     \label{eq:m3range1}
 \end{eqnarray}
where the approximation $\Delta m_{21}^2 \ll m_1 \approx m_2$ was applied. Therefore, the first correlation (Eq.~(\ref{eq:sumrule-A})) bound the mass $m_3$ within the range of $0.4\times10^{-3} \, {\rm eV^2} <m_3^2<1.3\times10^{-3} \,{\rm eV^2}$. 
Meanwhile, the second correlation (Eq.~(\ref{eq:sumrule-A2})) can be solved numerically for $m_3^2$  within the given range of $\theta_R$. The numerical results give $2.9\times10^{-3} \,{\rm eV^2}<m_3^2<3.1\times10^{-3} \,{\rm eV^2}$, which shows no overlap with the range predicted from Eq.~(\ref{eq:m3range1}).

Therefore, the free parameters in model A are not enough for fitting $\theta_{13}$, $\Delta m_{31}^2$, and $\Delta m_{32}^2$ simultaneously. Thus, Model A is over-constrained and excluded. As was mentioned in Section~\ref{sec:3.1}, modifying to higher modular weight of $\tau_\nu$ can fit the data with the price of introducing more free parameters, rendering model A less predictive and thus will not be considered any more in this work.


In Models B and C, since the charged lepton mass matrix is diagonal, simple forms can be written down for the mixing angle $\theta_{23}$ and Dirac-type CP-violating phase using the above parameterization~\cite{King:2019vhv}:
\begin{align}
    \tan \theta_{23} =& \left| \frac{\cos \theta_R +\sqrt{\frac23} e^{i(\alpha_1-\alpha_2)} \sin \theta_R }{\cos \theta_R -\sqrt{\frac23} e^{i(\alpha_1-\alpha_2)} \sin \theta_R } \right| \, ,     \label{eq:observables1.3} \\
    \delta =& {\rm arg}[(5(\cos 2\theta_R+1)\cos(\alpha_1-\alpha_2) \nonumber \\
    &-i( \cos 2\theta_R+5)\sin(\alpha_1-\alpha_2)] \, .
    \label{eq:observables1}
\end{align}
which recover the sum rules between mixing angle and CP phase shown in Eq.~(\ref{eq:sumrule}).

The lightest eigenvalue of $M_\nu$ under this parameterization is given by $m_{\rm lightest} = m_1$ for NO, and $m_{\rm lightest} = m_3$ for IO. 
Meanwhile, since $M_1$ is fixed by $\eta$ which is dependent on $\alpha$, $\beta$, $\gamma$ in our models, there should be a correlation between $M_1$, $M_2$ and $M_3$ that can be solved by diagonalizing the bottom-right $2\times 2$ block in Eq.~(\ref{eq:diag-mnu}). Using $\eta =2\alpha+\beta$ in Model B, we get:
\begin{align}
    \frac{1}{m_1}=&\frac{1}{|2\alpha+\beta|}=\left|\frac{2 e^{-2i\alpha_1} \cos^2 \theta_R + e^{-2i\alpha_2} \sin^2 \theta_R }{m_2} \right. \nonumber \\
    &+ \left. \frac{ e^{2i\alpha_1} \cos^2 \theta_R  + 2 e^{2i\alpha_2} \sin^2\theta_R }{m_3}\right| \, .
        \label{eq:observables2}
\end{align}
For Model C, the correlation is different because now we have $\eta = -\beta -2\gamma$:
\begin{align}
     \frac{1}{m_1} = & \frac{1}{| \beta +2\gamma|} = \left|\frac{ e^{-2i \alpha_2} \sin^2 \theta_R + e^{-i(\alpha_1+\alpha_2)} \sin 2\theta_R }{m_2} \right. \nonumber \\
     &+ \left.\frac{ e^{2i\alpha_1} \cos^2\theta_R - e^{i(\alpha_1+\alpha_2)} \sin2\theta_R }{m_3} \right| \, .
         \label{eq:observables3}
\end{align}

Furthermore, the effective neutrino mass parameter $m_{ee}$ in neutrino-less double beta decay experiments (see~\cite{Denton:2023hkx} for a review) can be solved 
for Model B
\begin{widetext}
\begin{align}
     m_{ee} = & m_0^2\left|{M_{R,B}^{-1}}_{(1,1)}\right|  =  m_0^2\left| \frac{2}{3 (2 \alpha + \beta)} + \frac{\beta }{3(\alpha \beta - \gamma^2)} \right| \nonumber \\
    = & \left| \frac{2m_2m_3}{3m_2( e^{2 i \alpha_1} \cos ^2\theta_R+2 e^{2 i \alpha_2}  \sin ^2\theta_R) + 3m_3 (e^{-2 i \alpha_2}  \sin ^2\theta_R+2e^{-2 i \alpha_1}  \cos ^2\theta_R)} \right.
    \nonumber \\
    &
    \left.+\frac13 {\left(  m_2 e^{2 i \alpha_1}  \cos ^2\theta_R + m_3 e^{-2 i \alpha_2} \sin ^2\theta_R \right)} \right| \, ,
        \label{eq:observables4}
\end{align}
and for Model C
\begin{align}
 m_{ee}= & m_0^2 \left|(M_{R,C}^{-1})_{(1,1)}\right| =  m_0^2 \left| \frac{2}{3(\beta+2\gamma)} -\frac{\beta}{3(\alpha\beta-\gamma^2)} \right|  \nonumber \\
=& \left| \frac{2m_2m_3}{3  m_2(e^{i2\alpha_1}\cos^2\theta_R -e^{i (\alpha_1+\alpha_2)}\sin 2\theta_R )+3m_3(e^{-i2\alpha_2} \sin^2\theta_R  +e^{-i (\alpha_1+\alpha_2)} \sin 2\theta_R )}  \right. \nonumber \\
& - \left.\frac13 \left(m_2 e^{2i\alpha_1} \cos ^2\theta_R
+ m_3 e^{-2i\alpha_2} \sin^2\theta_R \right) \right| \, . 
    \label{eq:observables5}
\end{align}
\end{widetext}


\subsection{Numerical Results}
In this section, we perform a numerical analysis for the parameters of Models B and C by applying the analytical expressions given in Eqs.~(\ref{eq:observables1})-(\ref{eq:observables5}). Though being expressed originally by five independent free parameters in the set $(|\alpha|,|\beta|,|\gamma|,{\rm arg}(\alpha),{\rm arg}(\beta))$, it is more convenient to re-parameterize the predictions of neutrino oscillation observables using parameters $\alpha_1$, $\alpha_2$, $\theta_R$. The ranges for parameter $\theta_R$ can be found by applying Eqs.~(\ref{eq:observables1.1})-(\ref{eq:observables1.2}) with the 1$\sigma$ or 3$\sigma$ bounds of $\theta_{13}$ and $\theta_{12}$ from JUNO~\cite{JUNO:2025gmd} and NuFIT~\cite{Esteban:2024eli}. There are additional parameters $m_i$ $(i=1,2,3)$ left as predictions of our model. The parameters $m_2$ and $m_3$ can be reduced by substituting the experimental values of $\Delta m_{12}^2$ and $\Delta m_{31}^2$ into their definition in either the NO or the IO scenario, and then $m_1$ can be solved numerically from Eqs.~(\ref{eq:observables2}) or (\ref{eq:observables3}).

In total, fixing Eqs.~(\ref{eq:observables1})-(\ref{eq:observables5}) to the experimental observed values within a certain range of uncertainties can reduce the number of free parameters from 5 to 2, including the two phases $\alpha_1$ and $\alpha_2$.  
We can now perform our numerical computations by freely sampling the two parameters $\alpha_1$ and $\alpha_2$ within the ranges $(0,2\pi)$ to fit the remaining two observables, $\sin \theta_{23}$ and $\delta$, within their 1$\sigma$ and $3 \sigma$ ranges. 

We also note that the MSSM renormalization group running does not qualitatively modify the TM$_1$ analysis used here when $\tan\beta$ is small. The leading correction to $\theta_{12}$ is controlled by the tau-Yukawa contribution and becomes significant only for the large $\tan\beta$ regime~\cite{Antusch:2001vn,Antusch:2003kp,Zhou:2014sya}. For modest $\tan\beta$ and away from quasi-degenerate neutrino masses, the induced RG shifts remain small and are not expected to qualitatively affect the low-energy TM$_1$ correlations used in our fits.

\begin{figure*}[t]
    \centering
    \includegraphics[width=0.45\linewidth]{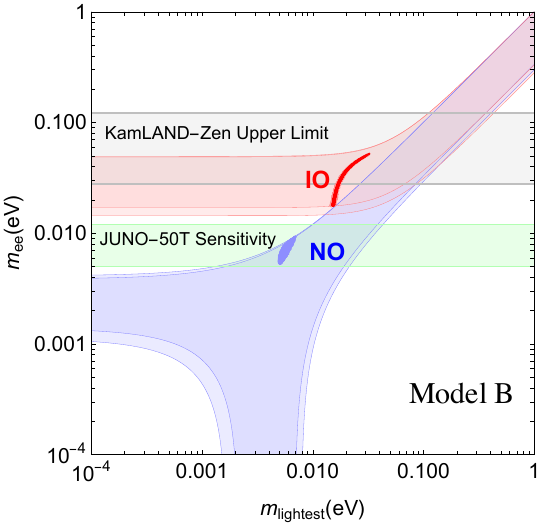}  \,
    \includegraphics[width=0.45\linewidth]{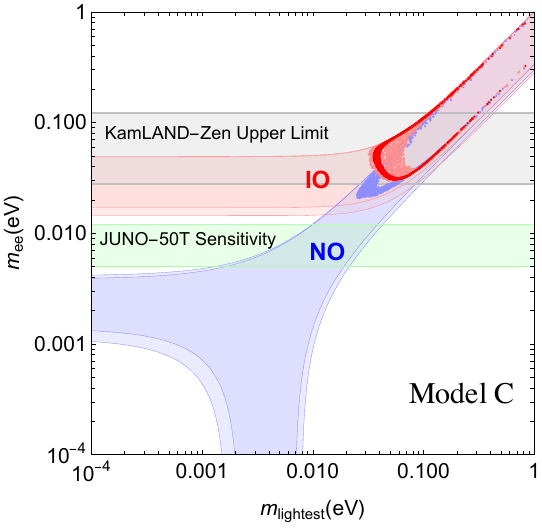} 
    \caption{$m_{\rm lightest}$ versus $m_{ee}$ predicted in Models B and C. Current experimental upper bounds and future sensitivities on $m_{ee}$ are shown for reference.} 
    \label{fig:modelB}
\end{figure*}

The numerical results for the prediction of the effective neutrino mass parameter $m_{ee}$ in neutrinoless double beta decay experiment with respect to the mass of the lightest neutrino $m_{\rm lightest}$ are shown in Figure \ref{fig:modelB} for Model B and Model C. These figures show the $1\sigma$ and $3\sigma$ ranges of oscillation parameters given by NuFIT 6.0~\cite{Esteban:2024eli} (in blue) and JUNO (in red) for the NO assuming $m_{\rm lightest} = m_1 $, and the IO assuming $m_{\rm lightest} = m_3 $. We also show the current upper limit of KamLAND-Zen experiment $(m_{ee})_{\rm upper}=0.028-0.122$ eV \cite{KamLAND-Zen:2024eml} (in gray) and future sensitivities in JUNO-50T $(m_{ee})_{\rm upper}=0.005-0.012$~eV \cite{Zhao:2016brs} (in green), in the figure. For the correlation between the mixing angles and the CP phase $\delta$, no additional model-specific features are predicted in Models B and C beyond the sum rules in Eq.~\eqref{eq:sumrule}, and thus a separate plot is not shown here.

\begin{figure*}[t]
    \centering
    \includegraphics[width=0.45\linewidth]{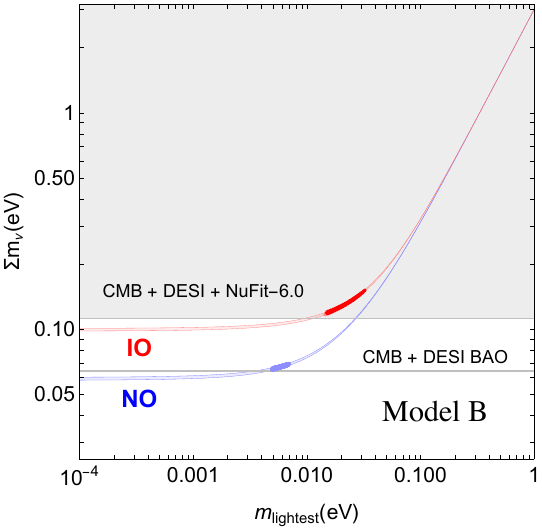}  \,
    \includegraphics[width=0.45\linewidth]{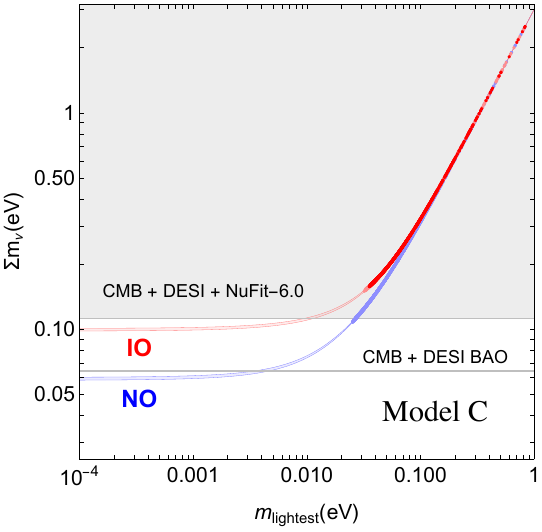} 
    \caption{Predicted $\sum m_\nu$ versus $m_{\rm lightest}$ in Models B and C for NO and IO. The lower and upper gray lines indicate, respectively, the 95\% C.L. upper bounds $\sum m_\nu < 0.0642~{\rm eV}$ from CMB + DESI DR2 BAO and $\sum m_\nu < 0.112~{\rm eV}$ from CMB + DESI DR2 BAO + NuFit-6.0 ~\cite{Planck:2018vyg,Tristram:2023haj,DESI:2025zgx,Elbers:2025vlz}.}
    \label{fig:summnu}
\end{figure*}

As a complementary cosmological test, we also compare the allowed Model B and Model C point sets with the recent upper bounds on the sum of neutrino masses, $\sum m_\nu < 0.0642$ eV (95\% C.L.) from CMB + DESI DR2 BAO and $ \sum m_\nu < 0.112$ eV (95\% C.L.) from CMB + DESI DR2 BAO + NuFit-6.0, as shown in Fig.~\ref{fig:summnu}~\cite{Planck:2018vyg,Tristram:2023haj,DESI:2025zgx,Elbers:2025vlz}. For Model B, the sampled NO solutions remain compatible with the CMB + DESI DR2 BAO + NuFit-6.0 bound, whereas the IO case is almost entirely above 0.112 eV, with only a small edge of the $3\sigma$ sample touching that bound. For Model C, the sampled IO solutions lie above $0.112~{\rm eV}$, while only a fraction of the NO solutions remain below this bound. Therefore, the cosmological comparison further disfavors the IO regions, especially in Model C, and also imposes pressure on the NO regions.

\section{Conclusions}\label{sec:5}

This paper investigates the landscape of models based on modular $S_4$ symmetry that predict the TM$_1$ pattern for neutrino mixing, and explores their parameter space compared with the latest, high-precision neutrino oscillation data from the JUNO experiment. We first show the consistency of TM$_1$ mixing with JUNO data, and then briefly review its realization by flavor symmetry. Starting from the permutation group $S_4$, we showed that when the residual symmetry is assigned to be $Z_3^T$ in the charged lepton sector and $Z_2^{SU}$ in the neutrino sector, the model will predict the TM$_1$ mixing pattern. The result holds in both the traditional flavor models and modular flavor models.

We then explore three possible approaches for concrete model building in modular symmetries, as shown in Figure \ref{fig:approaches}.
\begin{itemize}
    \item In Model A, in the charged lepton sector, the breaking of $S_4 \to Z_3^T$ implemented by adopting a flavon alignment 
    $\langle\varphi_l\rangle\sim(1,0,0)^T$, while in the neutrino sector, the residual $Z_2^{SU}$ was due to the modular field $\tau_\nu$ stabilizing at the fixed point $\langle \tau_\nu \rangle =\tau_{SU}$. 
    \item In Model B, the residual symmetry $Z_3^T$ in the lepton sector was realized by the modular field $\tau_l$ stabilizing at the fixed point $\langle \tau_l \rangle =\tau_{T}$, giving rise to a diagonalized charged lepton mass matrix, while the residual symmetry $Z_2^{SU}$ was enforced by the VEVs of two flavons aligned as $\langle\varphi_\nu\rangle\sim(1,1,1)^T$ and $\langle\varphi_\nu'\rangle\sim(0,1,-1)^T$. 
    \item In Model C, the modular symmetry acts on both the charged lepton and the neutrino sectors, thus a flavon $\Phi$ is needed to break the modular $S_4^\ell \times S_4^\nu$ symmetry to a diagonal $S_4$ subgroup. The two modular fields are then fixed at $\langle \tau_l \rangle =\tau_{T}$ and $\langle \tau_\nu \rangle =\tau_{SU}$ to realize the TM$_1$ mixing.
\end{itemize} 
Each model has five free parameters to fit five observables, including two mass-squared differences and three mixing angles, while the Dirac CP phase is regarded as a prediction.
For each model, we parameterize the explicit Majorana neutrino mass matrix by Eq.~(\ref{eq:parameterization}), which is then used to derive analytical predictions and perform numerical fits to neutrino oscillation data.

Model A shows an over-restricted correlation between $\theta_{13}$ and neutrino masses and is thus excluded. However, it can be saved by arranging different modular weights to right-handed neutrinos with the price of more free parameters. We perform numerical scans for Models B and C using the analytical expressions derived in Eqs.~(\ref{eq:observables1.1})-(\ref{eq:observables1.2}) and Eqs.~(\ref{eq:observables1.3})-(\ref{eq:observables5}). Parameter regions consistent with the latest JUNO and NuFIT constraints are explored, and the predicted correlations between $m_{\rm lightest}$ and $m_{ee}$ are shown in Figure~\ref{fig:modelB}. A complementary comparison with the cosmological baseline $\sum m_\nu  < 0.112$ eV further disfavors the IO regions, especially in Model C, while the NO region of Model B remains allowed. Overall, Model B provides the best fit among the three constructions considered here.


\bigskip 

\noindent {\bf Acknowledgements:} 
\noindent This work is supported by National Natural Science Foundation of China (NSFC) under Grants Nos. 12205064, 12347103, and Zhejiang Provincial Natural Science Foundation of China under Grant No. LDQ24A050002.


\appendix

\section{Vacuum alignment from driving fields}
\label{app:vacuum-alignment}
In this appendix, we give an explicit driving-field realization of the flavon alignments assumed in Models A and B. The construction follows the standard supersymmetric alignment mechanism used in conventional $S_4$ flavor models~\cite{Bazzocchi:2009pv,Ding:2009iy}. We introduce driving fields with vanishing VEVs and assign the $U(1)_R$ charges such that matter fields have $R=1$, flavons and Higgses have $R=0$, and driving fields have $R=2$. The driving superpotential is then linear in the driving fields, and the desired alignments follow from the corresponding $F$-term equations. We only present one possible realization; it is not meant to be a unique UV completion of the flavon sector.

\subsection*{Vacuum Alignment in Model A}
For Model A, we introduce the driving fields $\eta^0\sim{\bf 2}$, $f_1^0\sim{\bf 1}'$, $f_2^0\sim{\bf 1}'$. One possible set of auxiliary charges is given in Table~\ref{tab:alignment-A}. Modular forms are taken to be neutral under the auxiliary $Z_5$ symmetry.

\begin{table}[h]
\centering
\renewcommand{\arraystretch}{1.2}
\begin{tabular}{|c|cccccc|}
\hline
 & $\varphi_l$ & $\eta$ & $\xi$ & $\eta^0$ & $f_1^0$ & $f_2^0$ \\
\hline
$S_4$ & ${\bf 3}$ & ${\bf 2}$ & ${\bf 1}'$ & ${\bf 2}$ & ${\bf 1}'$ & ${\bf 1}'$ \\
$2k$ & $+2$ & $+2$ & $+6$ & $-4$ & $-6$ & $-6$ \\
$Z_5$ & $1$ & $3$ & $2$ & $3$ & $4$ & $3$ \\
\hline 
\end{tabular}
\caption{Transformation properties of the flavons and driving fields used for the vacuum alignment in Model A.}
\label{tab:alignment-A}
\end{table}

The matter-sector charges may be chosen as
\begin{align}
Z_5(L)=1,\quad Z_5(e^c)=2,\quad Z_5(\nu^c)=4,
\end{align}
with $\mu^c,\tau^c,H_u,H_d$ neutral. This assignment keeps the operators in Eq.~(\ref{eq:WmodelA}) as the leading terms and forbids lower-order competing terms.

The leading driving superpotential can be written as
\begin{align}
W_A^{\rm drv} &=   \mu_1  f_2^0\xi + \mu_2 M_{{\bf 3}'}^{(4)}(\tau_{SU})f_1^0\varphi_l + g_1 \eta^0(\varphi_l\varphi_l)_{\bf 2} 
\nonumber \\
& + g_2 Y_{\bf 2}^{(2)}(\tau_{SU})f_1^0(\eta\eta)_{\bf 2} + g_3 Y_{\bf 2}^{(2)}(\tau_{SU})f_2^0(\varphi_l\varphi_l)_{\bf 2}
\nonumber\\
&+ g_4 Y_{{\bf 3}'}^{(2)}(\tau_{SU})f_2^0(\varphi_l\varphi_l)_{\bf 3}
.
\label{eq:WA-driving}
\end{align}
where $\mu$'s and $g$'s are the usual quadratic and trilinear couplings, while $M_{\bf 3'}^{(4)}$ and $Y_{\bf r}^{(2k)}$ denote the modular Mass and modular Yukawa couplings with modular weight $2k$. In particular, $M_{\bf 3'}^{(4)}(\tau_{SU})\propto(2,-1,-1)^T$. 

Writing
\begin{align}
\varphi_l=(\varphi_1,\varphi_2,\varphi_3)^T,\qquad
\eta=(\eta_1,\eta_2)^T,
\end{align}
and expanding the superpotential using the CG coefficients, the vanishing $F$-term of $\eta^0$ gives
\begin{align}
\varphi_3^2+2\varphi_1\varphi_2&=0,\nonumber \\
\varphi_2^2+2\varphi_1\varphi_3&=0.
\end{align}
These equations admit the solution
\begin{equation}
\langle\varphi_l\rangle=v_T(1,0,0)^T .
\end{equation}
With this solution, the F-terms for the other two driving fields reduce to
\begin{align}
-g_2(\eta_1^2+\eta_2^2)+2\mu_2 v_T&=0,\\
\mu_1 \xi+2g_4v_T^2&=0,
\end{align}
where the modular forms have been expanded using eqs.~(\ref{eq:mf}). Therefore, the constraints on vacuum are 
\begin{equation}
\eta_1^2+\eta_2^2=\frac{2\mu_2}{g_2}v_T ,\qquad
\langle\xi\rangle=-\frac{2g_4}{\mu_1} v_T^2 .
\end{equation}
Thus the required alignment
\begin{equation}
\langle\varphi_l\rangle=v_T(1,0,0)^T ,\quad
\langle\eta\rangle=(v_{\eta_1},v_{\eta_2})^T,\quad
\langle\xi\rangle\neq0
\end{equation}
can be obtained. The direction of $\eta$ is not fixed at this order, which is sufficient for the phenomenological charged-lepton structure used in Model A. Other stationary points related by the flavor symmetry or lifted by soft terms may exist, as in the standard driving-field construction.

\subsection*{Vacuum Alignment in Model B}
For Model B, we denote
\begin{align}
\varphi_\nu=(\varphi_1,\varphi_2,\varphi_3)^T,\qquad
\varphi_\nu'=(\chi_1,\chi_2,\chi_3)^T \, ,
\end{align}
A possible realization for its vacuum alignment can be obtained by introducing one triplet driving field $\varphi^0\sim{\bf 3}$ and four singlet driving fields $f_i^0\sim{\bf 1}'$ $(i=1,\ldots,4)$. A possible minimal auxiliary symmetry is $Z_5$, with charges shown in Table~\ref{tab:alignment-B}. 
\begin{table}[h]
\centering
\renewcommand{\arraystretch}{1.2}
\begin{tabular}{|c|ccccccccc|}
\hline
 & $\varphi_\nu$ & $\varphi_\nu'$ & $\eta$ & $\xi$ & $\varphi^0$ & $f_1^0$ & $f_2^0$ & $f_3^0$ & $f_4^0$ \\
\hline
$S_4$ & ${\bf 3}$ & ${\bf 3}'$ & ${\bf 1}'$ & ${\bf 1}'$ & ${\bf 3}$ & ${\bf 1}'$ & ${\bf 1}'$ & ${\bf 1}'$ & ${\bf 1}'$ \\
$2k$ & $+4$ & $+3$ & $+1$ & $+2$ & $-8$ & $-7$ & $-9$ & $-6$ & $-8$ \\
$Z_5$ & $4$ & $1$ & $3$ & $2$ & $2$ & $0$ & $4$ & $1$ & $3$ \\
\hline
\end{tabular}
\caption{Transformation properties of the flavons and driving fields used for the vacuum alignment in Model B.}
\label{tab:alignment-B}
\end{table}

With $e^c,\mu^c,\tau^c$ neutral, the matter-sector charges may be chosen as
\begin{align}
Z_5(L)=1,\ Z_5(H_u)=1,\  Z_5(H_d)=-1,\  Z_5(\nu^c)=3 \, .
\end{align}

With these assignments, the leading driving superpotential may be written as
\begin{align}
W_B^{\rm drv} & = h_1 \varphi^0(\varphi_\nu\varphi_\nu)_{\bf 3} 
+h_2 f_1^0(\varphi_\nu\varphi_\nu')_{{\bf 1}'}
\nonumber\\
& + h_3 M_{\bf 3}^{(6)}(\tau_T) f_2^0  \varphi_\nu'  +
h_4  M_{{\bf 3}'}^{(2)}(\tau_T)f_3^0\varphi_\nu 
\nonumber\\
&+h_5 Y_{{\bf 3}'}^{(2)}(\tau_T)f_3^0 \left(\varphi_\nu'  \eta\right)_{\bf 3} + 
h_6 Y_{\bf 2}^{(2)}(\tau_T)f_4^0(\varphi_\nu'\varphi_\nu')_{\bf 2}
\nonumber \\
&
+h_7 Y_{{\bf 3}'}^{(2)}(\tau_T)f_4^0(\varphi_\nu'\varphi_\nu')_{\bf 3}
+h_8 M_{\bf 1}^{(6)}(\tau_T)f_4^0\xi
 .
\label{eq:WB-driving}
\end{align}
The last line involves $\varphi_\nu'\varphi_\nu'$, not $\varphi_\nu\varphi_\nu'$. This is required both by the auxiliary charge assignment and by the $F$-term equations below.

The $F$-term equations from $\varphi^0$ give
\begin{align}
\varphi_1^2-\varphi_2\varphi_3&=0,\nonumber \\
\varphi_2^2-\varphi_1\varphi_3&=0,\nonumber \\
\varphi_3^2-\varphi_1\varphi_2&=0.
\end{align}
For nonzero $\varphi_1$, these equations admit
\begin{equation}
\langle\varphi_\nu\rangle=v_{SU}(1,1,1)^T,
\label{eq:vevvarphinu}
\end{equation}
up to the usual degenerate solutions related by the flavor symmetry.

The $F$-term equations from $f_1^0$ and $f_2^0$ give
\begin{align}
v_{SU}\left(\chi_1+\chi_3+\chi_2\right) &=0 \, , \nonumber \\
\chi_1 & = 0  \, ,
\end{align}
where eq.~(\ref{eq:vevvarphinu}) has been substituted and, therefore,
\begin{equation}
\langle\varphi_\nu'\rangle=v'_{SU}(0,1,-1)^T.
\label{eq:vevvarphinup}
\end{equation}

The remaining two F-term equations fix the VEVs of the flavon singlets $\eta$ and $\xi$. After absorbing nonzero modular-form components into the coefficients, they may be written as
\begin{align}
 h_4 \varphi_3+h_5\eta\chi_3&=0 \, , \nonumber \\
 (h_6+2h_7)\chi_2^2 +2(h_6-h_7)\chi_1\chi_3 +h_8 \xi&=0 \, .
\end{align}
Substituting eqs.~(\ref{eq:vevvarphinu}) and (\ref{eq:vevvarphinup}), the above equation reduces to
\begin{align}
\langle\eta\rangle
 =\frac{h_4v_{SU}}{h_5v'_{SU}} \neq 0 \, , \,
\langle\xi\rangle
=-\frac{h_6+2h_7}{h_8 }(v'_{SU})^2 \neq 0 \, ,
\end{align}
which exactly reproduces the flavon vacuum alignments assumed at the beginning of Model B.



\end{document}